\documentclass[12pt, draftclsnofoot, onecolumn]{IEEEtran}

\usepackage{amsmath}
\usepackage{amssymb}
\usepackage{amsfonts}
\usepackage{graphicx}
\usepackage{epsfig}
\usepackage{subfigure}
\usepackage{psfrag}
\usepackage{cite}
\usepackage{latexsym}
\usepackage{url}
\usepackage{color}
\usepackage{multirow}
\usepackage{mathtools}
\usepackage{bm}
\usepackage{booktabs}
\usepackage{algorithm}
\usepackage{algpseudocode}

\usepackage{verbatim}
\usepackage{indentfirst}
\usepackage{hyperref}
\usepackage{multirow}

\graphicspath{{fig/}}
\PassOptionsToPackage{bookmarks={false}}{hyperref}

\IEEEoverridecommandlockouts

\newtheorem{remark}{\underline{Remark}}[section]
\newcommand{\mv}[1]{\mbox{\boldmath{$ #1 $}}}
\renewcommand{\algorithmicrequire}{\textbf{Input:}$~$}
\renewcommand{\algorithmicensure}{\textbf{Output:}$~$}

\begin{document}
\title{Semantic Communications for Image Recovery and Classification via Deep Joint Source and Channel Coding}
\author{Zhonghao Lyu, \emph{Graduate Student Member, IEEE,} Guangxu Zhu, \emph{Member, IEEE,} Jie Xu, \emph{Senior Member, IEEE,} Bo Ai, \emph{Fellow, IEEE,}\\ and Shuguang Cui, \emph{Fellow, IEEE}\\
\thanks{Part of this paper has been submitted to IEEE Global Communications Conference (GLOBECOM) 2023 \cite{ZL2023}.}
\thanks{Z. Lyu and J. Xu are with the School of Science and Engineering (SSE) and the Future Network of Intelligence Institute (FNii), The Chinese University of Hong Kong (Shenzhen), Shenzhen, China (e-mail: zhonghaolyu@link.cuhk.edu.cn, xujie@cuhk.edu.cn). }
\thanks{
G. Zhu is with Shenzhen Research Institute of Big Data, Shenzhen, China  (e-mail: gxzhu@sribd.cn). G. Zhu and J. Xu are corresponding authors.}
\thanks{B. Ai is with the State Key Laboratory of Rail Traffic Control and Safety, Beijing Jiaotong University, Beijing, China, and with Peng Cheng Laboratory, Shenzhen, China. He is also affiliated with Henan Joint International Research Laboratory of Intelligent Networking and Data Analysis, Zhengzhou University, Zhengzhou, China (e-mail: boai@bjtu.edu.cn).}
\thanks{S. Cui is with the SSE and FNii, The Chinese University of Hong Kong (Shenzhen), and Shenzhen Research Institute
of Big Data, Shenzhen, China. He is also affiliated with Peng Cheng Laboratory, Shenzhen, China (e-mail:
shuguangcui@cuhk.edu.cn).}
}
\maketitle

\vspace{-10pt}
\begin{abstract}
With the recent advancements in edge artificial intelligence (AI), future sixth-generation (6G) networks need to support new AI tasks such as classification and clustering  apart from data recovery. Motivated by the success of deep learning, the semantic-aware and task-oriented communications with deep joint source and channel coding (JSCC) have emerged as new paradigm shifts in 6G from the conventional data-oriented communications with separate source and channel coding (SSCC). However, most existing works focused on the deep JSCC designs for one task of data recovery or AI task execution independently, which cannot be transferred to other unintended tasks. Differently, this paper investigates the JSCC semantic communications to support multi-task services, by performing the image data recovery and classification task execution simultaneously. First, we propose a new end-to-end deep JSCC framework by unifying the coding rate reduction maximization and the mean square error (MSE) minimization in the loss function. Here, the coding rate reduction maximization facilitates the learning of discriminative features for enabling to perform classification tasks directly in the feature space, and the MSE minimization helps the learning of informative features for high-quality image data recovery. Next, to further improve the robustness against variational wireless channels, we propose a new gated deep JSCC design, in which a gated net is incorporated for adaptively pruning the output features to adjust their dimensions based on channel conditions. Finally, we present extensive numerical experiments to validate the performance of our proposed deep JSCC designs as compared to various benchmark schemes. It is shown that our proposed designs simultaneously provide efficient multi-task services, and the proposed gated deep JSCC framework efficiently reduces the communication overhead with only marginal performance loss. It is also shown that performing the classification task on the feature space via coding rate reduction maximization is able to better defend the label corruption than the traditional label-fitting methods.
\end{abstract}
\vspace{-15pt}

\section{Introduction}
Recently, the unprecedented development of modern communication and artificial intelligence (AI) technologies have witnessed the successful deployment of various AI applications at the network edge, such as auto-driving, virtual reality (VR), and Metaverse, which breed a forward-looking vision of a more intelligent and collaborative human society\cite{DCNguyen2022,KB2022,GZhu2023}. With more frequent and intelligent interaction between humans, machines, and the environment for various task goals, the conventional communication systems with separate source and channel coding (SSCC) designs regardless of specific application tasks may not meet the stringent service requirements of extremely high data rates and ultra-low latency \cite{Strinati2021}. Fortunately, semantic communications \cite{SMa2023} (and task-oriented or goal-oriented communications \cite{DGunduzZQin2023,Strinati2021,PLiu2023}) could enjoy lower communication overhead as well as being more robust to terrible channel conditions for high-reliable communications, via extracting and transmitting semantic-aware information in a task-oriented manner.  In such context, semantic communications provide novel communication design paradigm shifts beyond traditional Shannon's model \cite{Shannon} by jointly designing the wireless transmission together with specific application tasks to fully unlock the potential of future wireless networks.

Recently, the rapid development of deep learning (DL) has inspired sparkled research interest in deep joint source and channel coding (JSCC) to enable semantic communications. In deep JSCC, the semantic-aware information is extracted by deep neural networks (DNNs), which is highly task-specific \cite{HZhang2022}. Thus, it is crucial to design proper network structures and loss functions according to specific tasks. In general, the tasks could be characterized as two core categories, namely \emph{data recovery} and \emph{AI task execution}. 

For \emph{data recovery}, existing works have designed different deep JSCC semantic communication frameworks targeting different types of transmission data, e.g., image \cite{Bourtsoulatze2019,JYan2021,DHuang2023,JDai2022}, text \cite{HXie2021}, speech \cite{ZWen2021}, and sensing data \cite{HDu2023}. For example, the authors in \cite{Bourtsoulatze2019} considered an end-to-end deep JSCC framework for  wireless image transmission, where they extracted the semantic features and recovered the original images based on the design principle of mean square error (MSE) for pixel-level image reconstruction. Apart from MSE, the authors in \cite{JYan2021} proposed a semantic communication framework based on an autoencoder to transmit images. The autoencoder is trained by using the structural similarity index matrix (SSIM), which is another widely used design criterion in computer vision \cite{ZWang2004} for measuring the similarity of two images in terms of luminance, contrast, and
structure. Moreover, there are some other works concerning  wireless image transmission based on different loss function designs such as rate-semantic-perceptual \cite{DHuang2023} and nonlinear transform coding \cite{JDai2022}. However, the proposed frameworks in above works only focused on data recovery, which fail to serve on-growing needs of executing AI tasks in future wireless networks.

For \emph{AI task execution}, main works have focused on extracting task-relevant features to support successful AI task execution (such as classification and clustering). For example, the authors in \cite{Jankowski2020} proposed an image classification oriented wireless transmission framework for wireless person re-identification based on the cross-entropy loss. Moreover, inspired by information bottleneck, image classification tasks were considered in \cite{Shao2022} and \cite{Shao2023} via extracting minimum sufficient features for the concerned tasks through task-oriented deep JSCC frameworks. Furthermore, considering multi-modal data, the authors in \cite{HXie2022} extracted features from correlated multi-modal data to perform visual question answering tasks. However, the deep JSCC frameworks were designed only for AI task execution in above works, where the latent semantic features are not helpful for recovering the original data. In addition, the design criteria  in existing semantic communication frameworks for AI task execution mainly focused on end-to-end label fitting, which is highly label-dependent and may suffer from mislabelling (label corruption).

With the analysis above, the proposed deep JSCC frameworks in existing works were dedicatedly designed for handling only one task (either data recovery or AI task execution). When performing the tasks different from the model being designed for, it will not work due to the model mismatch and/or the inappropriate training pipeline (e.g., the loss function that the model is trained with).  However, in practice, many applications require performing data recovery and AI task execution simultaneously. For example, in some VR applications, environment reconstruction and object classification need to be jointly performed \cite{BOmarali2022}. To this end, it is highly desired to design a deep JSCC framework for semantic communications to provide data recovery and AI task execution services simultaneously, which motivates our investigation in this work. Moreover, to tackle the potential label corruption issue, different from existing works, we invoke the idea of contrastive learning \cite{AJaiswal2021}, specifically, the coding rate reduction method\cite{YMa2007,YYu2020,XDai2021}, to learn linear discriminative representations for multi-class data. In such a way, we could perform the downstream classification task directly in the feature space, which enjoys the benefits of lower complexity and higher robustness \cite{YXue2022}.

In this work, we consider to provide multi-task services in wireless networks by performing the image data recovery and classification task execution at the same time. Specifically, we invoke the semantic communication architecture with deep JSCC for data transmission, where the transmitter extracts and transmits the semantic-aware features of original images via the deep JSCC encoder in a task-oriented manner over a point-to-point wireless channel. Once the receiver receives the extracted features, it performs the image classification task directly based on the received features, and recovers original images through the deep JSCC decoder simultaneously. Our main results are summarized as follows.
\begin{itemize}
  \item First, we design a novel deep JSCC framework enabling end-to-end training via unifying coding rate reduction maximization and MSE minimization in the loss function. On one hand, with coding rate reduction maximization, the learned features are discriminative enough to support image classification task directly in the feature space. On the other hand, with MSE minimization, the learned features are informative enough to be captured by the decoder for high-quality data recovery. 
 \item Second, to improve the model robustness against variational channels as well as further reduce the communication overhead, we propose a gated deep JSCC framework with a gated net for adaptively activating output features according to variational channel conditions. The whole network is trained with domain randomization by randomly perturbing the channel condition to a wide range of signal-to-noise ratio (SNR) to make the learned deep JSCC framework  suitable to variational channel conditions.
  \item Finally, we present extensive numerical results to validate the performance of our proposed designs versus benchmark schemes considering deep JSCC designs with MSE and SSIM as loss functions, as well as SSCC design with JPEG2000 as source coding and capacity-achieving channel coding. First, it is shown that our proposed deep JSCC framework efficiently provides image recovery and classification services simultaneously against channel impairments, while the JSCC benchmarks could only perform well on one task that they are dedicatedly designed for. Furthermore, our proposed design does not suffer from ``cliff effect'' as compared to the SSCC scheme. Second, it is observed that the proposed gated deep JSCC framework is more robust to variational channels as well as being capable of further reducing the communication overhead with only marginal performance loss. Third, performing the classification task on the feature space via our considered coding rate reduction maximization principle is able to defend label corruption as compared to the traditional label-fitting method.
\end{itemize}

The remainder of this paper is organized as follows. Section II introduces the system model. Section III discusses the proposed deep JSCC framework for image recovery and classification task execution. Section IV presents the gated deep JSCC framework training with domain randomization. Section V provides numerical results to demonstrate the efficiency of our proposed designs. Section VI concludes this paper.

{\emph {Notations}}: Boldface letters represent vectors (lower case) or matrices (upper case). For an arbitrary-sized matrix ${\boldsymbol A}$,  ${\boldsymbol A}^{\rm T}$ and ${\boldsymbol A}^{\rm H}$ denote its transpose and conjugate transpose, respectively. For a square
matrix ${\boldsymbol B}$, ${\rm tr}({\boldsymbol B})$  denotes its trace. ${\boldsymbol I}$ and $\mv{0}$ denote identity and all-zero matrices, respectively. $\mathbb{C}^{m}$ and $\mathbb{R}^{m}$ denote the spaces of $m$ dimensional complex and real vectors, respectively. $\|\cdot\|$ denotes the Euclidean norm of a complex vector. $\mathbb{E}( \cdot )$ denotes the statistical expectation. $\circ$ denotes the element-wise product. $\left\lfloor \cdot \right\rfloor$ denotes the operation of rounding down to the nearest integer.

\section{System Model}
We consider an edge AI system, in which  the transmitter transmits images to the receiver over a point-to-point wireless link to perform joint image recovery and classification tasks at the same time. 

First, we introduce the image source. Specifically, let $\boldsymbol{s} \in \mathbb{R}^{L \times W \times C}$ denote the input image  with $L$, $W$, and $C$ denoting the height, width, and the number of color maps of $\boldsymbol{s}$, respectively. Also, we denote $B=L \times W \times C$ as the number of pixels in one image.

Next, we consider the wireless transmission of the image source. First, the transmitter maps $\boldsymbol{s}$ to complex-valued symbols $\boldsymbol{x} \in \mathbb{C}^{b}$, with $b$ denoting the number of symbols for transmission. Similar as in the previous deep JSCC works \cite{Bourtsoulatze2019,JYan2021}, we refer to $b/B$ as the compression ratio to evaluate the degree of image compression. Moreover, $\boldsymbol{x}$ should satisfy the average power constraint
\begin{align}\label{power constraint}
	\frac{1}{b}\mathbb{E} \|\boldsymbol{x}\|^2 \le 1.
\end{align}
Then, the encoded signal ${ \boldsymbol{x}}$ are transmitted through the wireless channel with $h$ denoting the channel coefficient. In particular, we consider a narrow-band or frequency-flat block fading channel, where the channel is assumed to be constant during the transmission of one image and may change for the next image independently. Then the received signal $\boldsymbol{y} \in \mathbb{C}^{b}$ is represented as
\begin{align}\label{y}
\boldsymbol{y}=h{\boldsymbol{x}}+\boldsymbol{n},
\end{align}
where $\boldsymbol{n} \in \mathbb{C}^{b}$ denotes the independent identically distributed (i.i.d.) circularly symmetric complex Gaussian (CSCG) noise vector with average noise power $\sigma^2$, i.e., $\boldsymbol{n} \sim \mathcal{CN}(0,\sigma^2\boldsymbol{I})$.

Finally, we introduce the receiver processing for image recovery and classification task execution. First, the receiver performs channel equalization by multiplying $1/h$ on both sides of \eqref{y} to obtain 
 \begin{align}\label{y_equalization}
	\hat{\boldsymbol{y}}={\boldsymbol{x}}+\hat{\boldsymbol{n}},
\end{align}
 where $\hat{\boldsymbol{n}}={\boldsymbol{n}}/h$ denotes the equivalent noise. Next, the receiver performs the data recovery and classification tasks based on $\hat{\boldsymbol{y}}$. On one hand, the  receiver performs classification task directly on the feature space by inputting the obtained features $\hat{\boldsymbol{y}}$ into a  pragmatic function $\phi(\hat{\boldsymbol{y}})$ to obtain the classification result ${z}=\phi(\hat{\boldsymbol{y}})$.\footnote{The pragmatic function $\phi(\hat{\boldsymbol{y}})$ could be given by parameterized methods such as DNN, or non-parameterized methods such as nearest subspace classifier, K-NearestNeighbor (KNN).} On the other hand, the decoder of the receiver maps the obtained $\hat{\boldsymbol{y}}$ to the estimated reconstruction of the original transmitted image $\hat {\boldsymbol{s}} \in \mathbb{R}^{B}$.

Our goal is to extract and transmit task-relevant semantic information $\mv{x}$ of the original image $\mv{s}$ to minimize the communication overhead (in terms of the number of symbols $b$ to be transmitted), while guaranteeing the performance of image recovering and classification tasks. Specifically, we evaluate the performance of image recovery task by the peak signal-to-noise ratio (PSNR), and evaluate the performance of image classification task by the classification accuracy, which are defined as follows. 
\begin{itemize}
    \item {\bf PSNR}: PSNR is a metric to evaluate the similarity of two images, which is defined as
\begin{align}\label{PSNR equation}
{\rm{PSNR}}(\boldsymbol{s},\hat{ \boldsymbol{s}}) = 10{\log _{10}}\frac{{{{MAX}}{^2}}}{\frac{1}{B}{\| {{\boldsymbol{s}} - {{\hat{ \boldsymbol{s}}}}} \|} ^2},
\end{align}
where ${{MAX}}$ is the maximum possible value of the image pixels.
\item {\bf Classification accuracy}: Classification accuracy is defined as the ratio of correctly predicted images $\{j'\}$ among all the test images $\{n'\}$, i.e., ${\frac{J'}{N'}}\%$, with $J'$ and $N'$ denoting the number of samples in $\{j'\}$ and $\{n'\}$, respectively. 
\end{itemize}

However, most existing works \cite{Bourtsoulatze2019,JYan2021,DHuang2023,JDai2022,HXie2021,ZWen2021,HDu2023,Jankowski2020,Shao2022,Shao2023,HXie2022} cannot achieve the above goals. This is because these methods were dedicatedly designed for performing  either data recovery or AI task execution, which fail to transfer to other unintended tasks for multi-task services. To this end, we aim at designing a novel deep JSCC framework to achieve the above goal of providing multi-task services by performing image recovery and classification task execution at the same time. In the following, we first introduce the proposed deep JSCC framework to enable performing image recovery and classification task execution simultaneously in Section III. Then, to improve the robustness against variational channel conditions as well as further reduce the communication overhead, we present the gated deep JSCC framework with domain randomization in Section IV.

\section{Deep JSCC Framework for Image Recovery and Classification Task Execution}
In this section, we propose a new semantic-and-task-aware deep JSCC framework to enable performing image recovery and classification tasks at the same time. In the following, we present the network structure and loss function design, respectively.

\begin{figure}[h]
	\centering
	 \epsfxsize=1\linewidth
		\includegraphics[width=16.5cm]{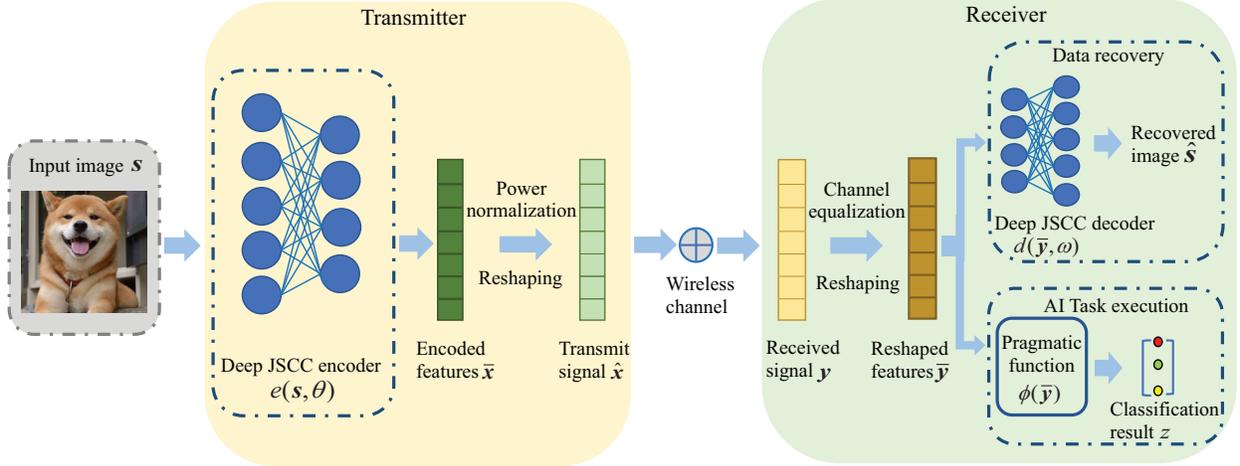}
	\caption{\label{model}Illustration of the proposed deep JSCC framework for image recovery and classification task execution.}
\end{figure}

\subsection{The Structure of Deep JSCC Framework}
The overall network structure is shown in Fig. \ref{model},  which includes two components, namely  the deep JSCC encoder and deep JSCC decoder. In the following, we discuss the deep JSCC encoder, wireless channel, and deep JSCC decoder, respectively.

First, we introduce the encoding process. Specifically, the deep JSCC encoder is parameterized by a DNN as $e(\cdot,\theta)$ with $\theta$ representing the parameters, and the output encoded semantic features $\bar{\mv{x}}\in \mathbb{R}^{2b}$ of $e(\cdot,\theta)$ are  denoted as $\bar{\boldsymbol{x}}=e(\boldsymbol{s},\theta)$. Then we combine (reshape) $\bar{\mv{x}}$ into $b$ complex-valued symbols to form the encoded signal $\boldsymbol{x}$. After encoding the task-relevant semantic signal $\boldsymbol{x}$ from the transmitted signal $\boldsymbol{s}$, we normalize it as
\begin{align}\label{nomarlized x}
\hat{\boldsymbol{x}}= p(\boldsymbol{x})=\sqrt d \frac{\boldsymbol{x}} { \sqrt {{\boldsymbol{x}^{\rm H}}\boldsymbol{x}}},
\end{align}
such that the transmit signal $\hat{\boldsymbol{x}}$ satisfies the average transmit power constraint in \eqref{power constraint}. 

Next, we consider the transmission of the encoded signal over the wireless channels after the encoding and normalization operations, i.e., $\boldsymbol{y}=h{\hat {\boldsymbol{x}}}+\boldsymbol{n}$, where the communication channels $h$ are integrated as non-trainable layers into the whole deep JSCC framework to enable end-to-end training \cite{HYe2020}. 

Finally, we discuss the processing of the received signal ${\boldsymbol{y}}$ at the receiver to perform image recovery and classification tasks simultaneously, which is different from semantic communication frameworks dedicatedly designed for only one task in previous works \cite{Bourtsoulatze2019,JYan2021,DHuang2023,JDai2022,HXie2021,ZWen2021,HDu2023,Jankowski2020,Shao2022,Shao2023,HXie2022}. First, the receiver applies channel equalization according to \eqref{y_equalization} to obtain $\hat{\boldsymbol{y}}={\hat{\boldsymbol{x}}}+\hat{\boldsymbol{n}}$. It is worth noting that, with channel 
equalization, we could implement the proposed deep JSCC framework on Rayleigh and Rician fading channels via directly transferring the model trained on the additive white gaussian noise (AWGN) channels with the corresponding SNR. Then the real and imaginary parts of $\hat{\boldsymbol{y}}$ are combined (reshaped) into $\bar{\boldsymbol{y}} \in \mathbb{R}^{2b}$ for further processing. Based on the obtained features $\bar{\boldsymbol{y}}$, we further perform image recovery  and classification tasks. On one hand, for the classification task, we directly conduct it on the feature space to obtain the classification result ${z}=\phi(\bar{\boldsymbol{y}})$. On the other hand, for the data recovery task, the decoder reconstructs the original image by $\hat{\boldsymbol{s}}=d(\bar{\boldsymbol{y}},\omega)$, where the decoding DNN is shown as $d(\cdot,\omega)$ with parameters $\omega$. The whole algorithm workflow for the proposed deep JSCC framework is summarized in Algorithm \ref{alg1}.

\begin{algorithm}[t]
	\caption{Algorithm workflow of the proposed deep JSCC framework}
	\label{alg1}
	\begin{algorithmic}[1] 
	\Statex *******************************{\it Training Phase}*******************************
	\State \algorithmicrequire A batch of images $\mv{S}$ to be transmitted.
	\State \algorithmicensure The trained deep JSCC framework with encoder $e(\cdot,\theta)$ and decoder $d(\cdot,\omega)$.
	\State {\bf Initialization:} Initialize the weights of $e(\cdot,\theta)$ and $d(\cdot,\omega)$ via normal initialization method.
	\For{each epoch}
	\Statex $\quad${\bf Transmitter:}
	\State $\quad$ Extract semantic features via $e(\mv{S},\theta)$.
	\State $\quad$ Perform power normalization, transmit the symbols over wireless channels.
	\Statex $\quad${\bf Receiver:}
	\State $\quad$ Perform channel equalization on the received signal $\mv{Y}$.
	\State $\quad$ Perform image recovery via $d(\bar{\boldsymbol{Y}},\omega)$ and image classification via $\phi(\bar{\mv{Y}})$. 
    \State Compute loss function value via \eqref{overall loss}, and train $e(\cdot,\theta)$ and $d(\cdot,\omega)$.
	\EndFor
	\Statex *******************************{\it Inference Phase}*******************************
	\State \algorithmicrequire An image $\mv{s}$ to be transmitted.
	\State \algorithmicensure The recovered image $\hat{\boldsymbol{s}}$ and the classification result ${z}$.
	\Statex {\bf Transmitter:}
	\State $\quad$ Deploy the learned $e(\cdot,\theta)$. Extract semantic features $\mv{x}$.
	\State $\quad$ Perform power normalization, and transmit the symbols $\hat{\boldsymbol{x}}$ in \eqref{nomarlized x} over wireless channels.
	\Statex {\bf Receiver:}
	\State $\quad$ Deploy the learned $d(\cdot,\omega)$. Perform channel equalization on the received signal $\mv{y}$.
	\State $\quad$ Perform image recovery via $d(\bar{\boldsymbol{y}},\omega)$ and image classification via $\phi(\bar{\mv{y}})$ 
	\end{algorithmic}
\end{algorithm}

\subsection{End-to-end Training based on Unified Loss Function}
In the following, we introduce the loss function design for the end-to-end training of the proposed deep JSCC framework for extracting task-relevant semantic features in a task-oriented manner. Specifically, to enable image recovery and classification task execution, the loss function design contains two parts. The first part is designed for  classification task execution  based on coding rate reduction maximization, and the second part is for data recovery based on MSE minimization. 

\begin{figure}[t]
	\centering
	\subfigure[]{
	\centering
	\includegraphics[width=0.362\linewidth]{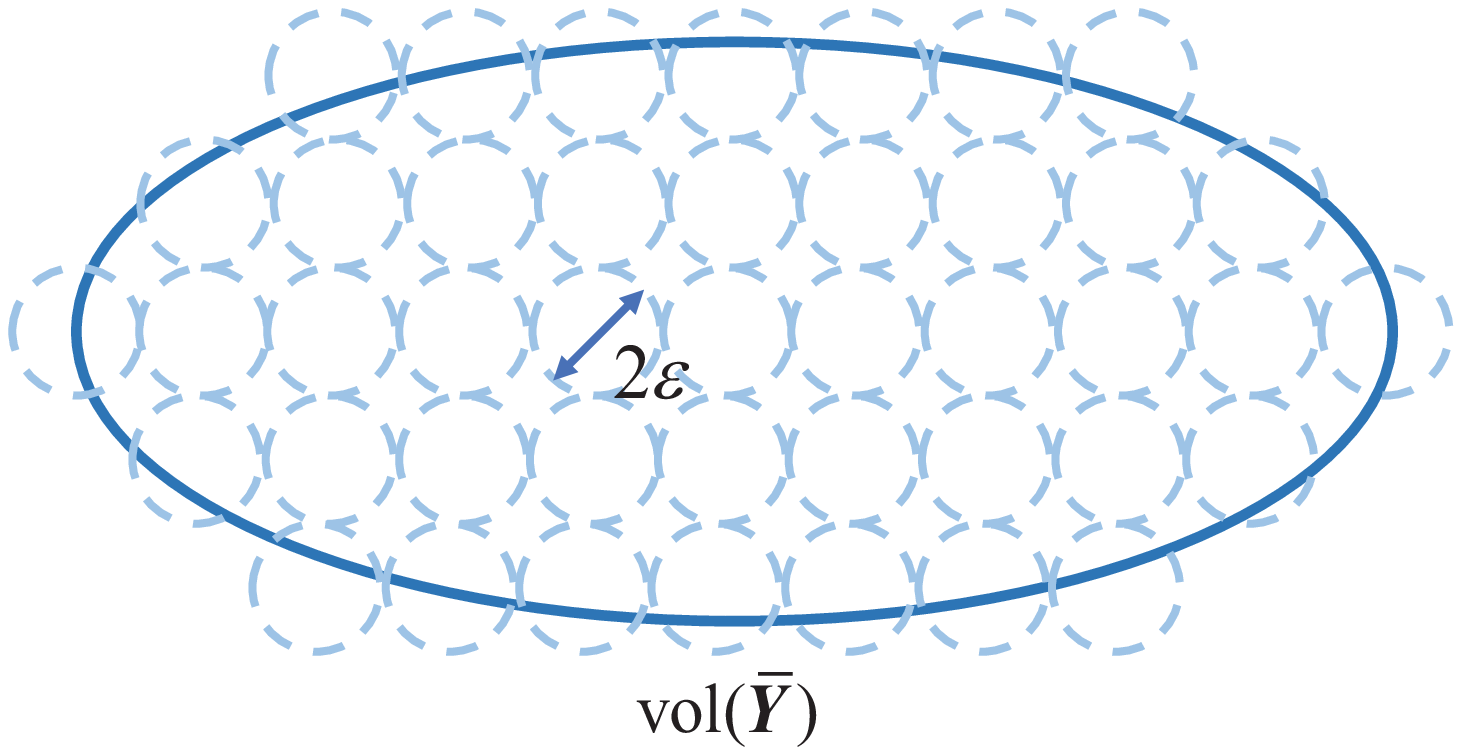}
	}%
	\subfigure[]{
	\centering
	\includegraphics[width=0.29\linewidth]{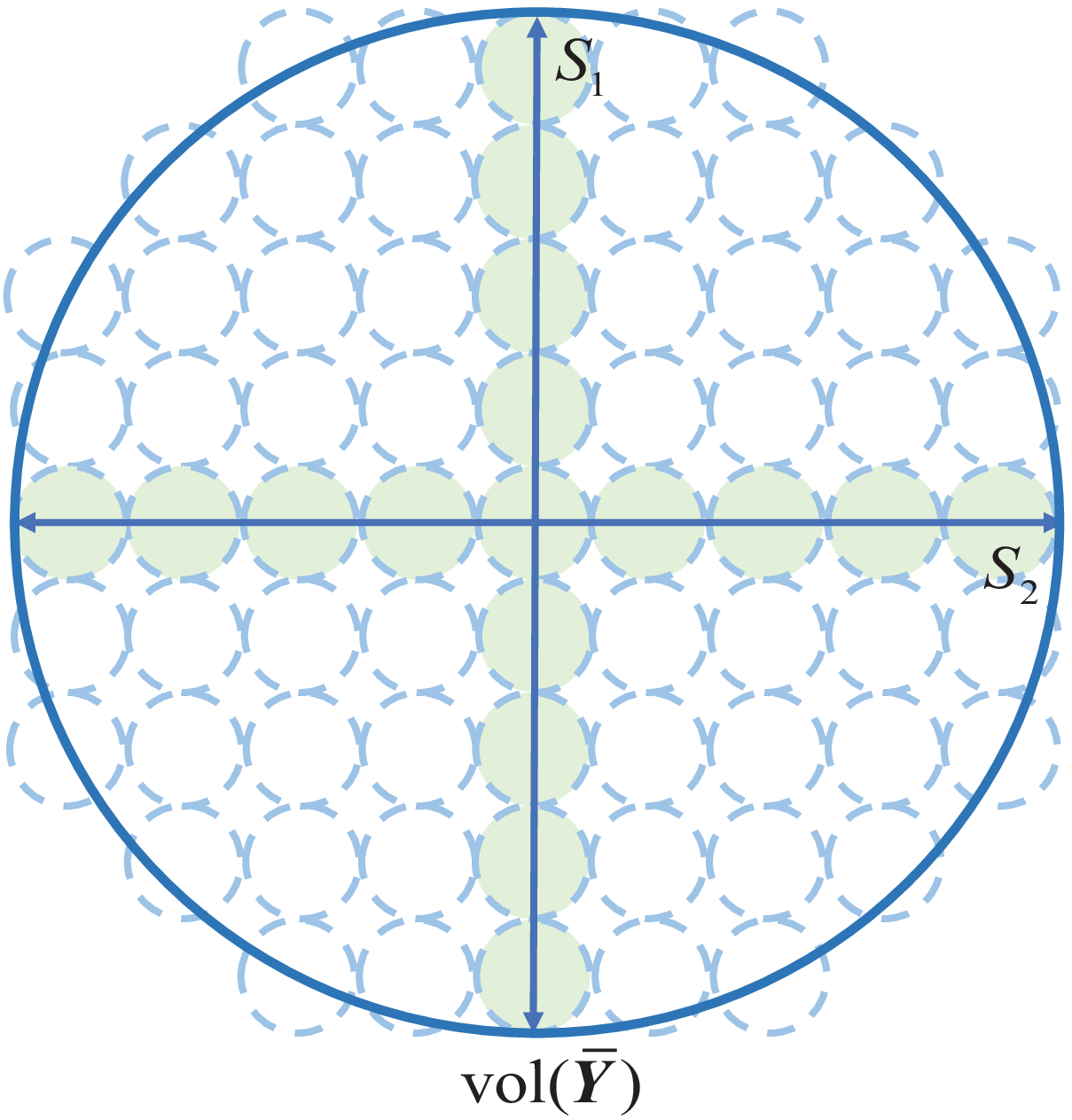}
	}%
	\subfigure[]{
	\centering
	\includegraphics[width=0.317\linewidth]{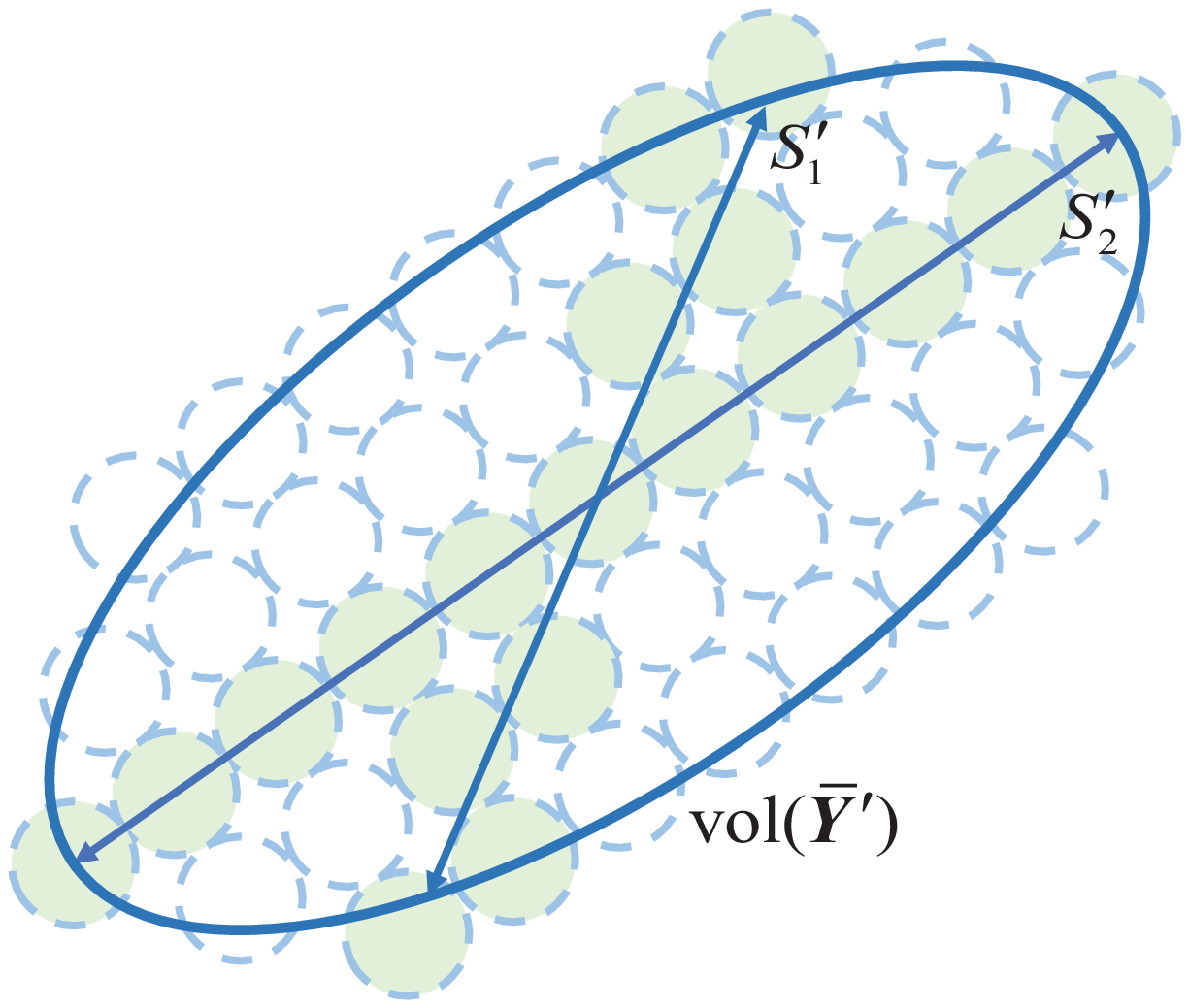}
	}%
	\centering
	\vspace{-10pt}
	\caption{Illustration of coding rate reduction from the viewpoint of sphere packing. (a) Using rate distortion theory to measure the volume ${\rm vol}(\bar{\boldsymbol{Y}})$  of the space spanned by $\bar{\boldsymbol{Y}}$. (b) $\&$ (c) Rate reduction comparison of two obtained features $\bar{\boldsymbol{Y}}$ and $\bar{\boldsymbol{Y}}'$, where $R(\bar{\boldsymbol{Y}} ,\varepsilon )$ in \eqref{rate distortion whole} indicates the total number of balls in the spanned space, ${R^d}(\bar{\boldsymbol{Y}}|\boldsymbol{\Pi} ,\varepsilon )$ in \eqref{rate distortion class} indicates the number of solid balls in subspaces $S_1$ and $S_2$ (or  $S_1'$ and $S_2'$), and $\Delta R(\bar{\boldsymbol{Y}}|\boldsymbol{\Pi} ,\varepsilon)$ in \eqref{rate reduction} is the number of hollow balls, which denotes the rate reduction.}
	\label{sphere packing}
\end{figure}

For ease of exposition, we first denote $\boldsymbol{S}=[\boldsymbol{s}_1,\ldots,\boldsymbol{s}_N]$ and $\hat{\boldsymbol{S}}=[\hat{\boldsymbol{s}}_1,\ldots,\hat{\boldsymbol{s}}_N]$ as sets of $N$ original and recovered data samples from $J$ classes for transmission, respectively. Also, we denote the set of obtained feature vectors as $\bar{\boldsymbol{Y}}=[\bar{\boldsymbol{y}}_1,\ldots,\bar{\boldsymbol{y}}_N]$. In the following, we present the coding rate maximization and the MSE minimization, respectively. 

\subsubsection{Coding Rate Reduction Maximization}
Coding rate reduction maximization is invoked to extract latent semantic features of $\boldsymbol{s}$ with clear subspace structures to support efficient implementation of down-stream classification tasks directly in the feature space. In the following, we first discuss the measurement of the volume of the space spanned by the obtained features $\bar{\boldsymbol{Y}}$ via rate distortion theory from information theory \cite{Tcover1991}. Then we discuss the coding rate reduction maximization principle for discriminative feature extraction.

First, we introduce how to measure the volume (the ``compactness'') ${\rm vol}(\bar{\boldsymbol{Y}})$ of the space spanned by the obtained features $\bar{\boldsymbol{Y}}$ via rate distortion theory, which is the foundation of the coding rate reduction principle. Generally, it is non-trivial to find an efficient measurement for the volume of the space spanned by a set of random variables from limited samples. Fortunately, in information theory, rate distortion \cite{Tcover1991} provides a method to tackle such problem. Specifically, according to nonasymptotic rate distortion theory \cite{YMa2007}, the volume of the space spanned by  $\bar{\boldsymbol{Y}}$ is measured mediately by 
\begin{align}\label{rate distortion whole}
	R(\bar{\boldsymbol{Y}} ,\varepsilon )= \frac{1}{2}\log \det \left( \boldsymbol{I} +  \frac{2b}{N {\varepsilon^2}} \bar{\boldsymbol{Y}}{\bar{\boldsymbol{Y}}^{\rm T}} \right),
	\end{align}
where $\varepsilon$ is a hyper-parameter representing the prescribed coding distortion. We explain it from the viewpoint of sphere packing to gain easy understanding, as shown in Fig. \ref{sphere packing}(a). Intuitively, the number of the spheres in Fig. \ref{sphere packing}(a) roughly represents the number of vectors that could be identified and coded in the space spanned by $\bar{\boldsymbol{Y}}$ with the coding accuracy up to $\epsilon^2$. To this end, the volume ${\rm vol}(\bar{\boldsymbol{Y}})$ of the space spanned by $\bar{\boldsymbol{Y}}$ is measured mediately by the approximated number of spheres $R(\bar{\boldsymbol{Y}} ,\varepsilon )$ in \eqref{rate distortion whole} with radius $\epsilon$ that could be packed in ${\rm vol}(\bar{\boldsymbol{Y}})$.

Then, we evaluate the rate distortion of the obtained features from each separate class. Specifically, we first encode a set of diagonal matrixes $\boldsymbol{\Pi}=\{{\boldsymbol{\Pi}}^j \in \mathbb{R}^{N\times N}\}$ to indicate pairwise relationships between the samples and the classes, where $\boldsymbol{\Pi}^j_{n,n}=1$ represents that sample $n$ belongs to class $j$ (otherwise, $\boldsymbol{\Pi}^j_{n,n}=0$). Then we evaluate the volume of the space spanned by  $\bar{\boldsymbol{Y}}$ concerning the belonging classes of the samples as
\begin{align}\label{rate distortion class}
{R^d}(\bar{\boldsymbol{Y}}|\boldsymbol{\Pi} ,\varepsilon )= \sum\nolimits_{j = 1}^J \frac{{\rm tr}(\boldsymbol{\Pi}^j)}{2N}\log \det (\boldsymbol{I} + \frac{2b}{{\rm tr}(\boldsymbol{\Pi}^j){\epsilon^2}}\bar{\boldsymbol{Y}}{\boldsymbol{\Pi} ^j}{\bar{\boldsymbol{Y}}^{\rm T}}).
\end{align}

With \eqref{rate distortion whole} and \eqref{rate distortion class}, we are now ready to invoke the rate reduction as loss function for classification tasks. To enable performing classification tasks directly in the feature space, we highly expect the extracted features to characterize the intrinsic statistical properties of the raw data \cite{YMa2007,YYu2020}. Specifically, it is highly desirable for the latent semantic features of $\boldsymbol{s}$ to meet the following properties: 
\begin{itemize}
    \item \textbf{Cross-class discrimination}: The features of samples across different classes are discriminative and uncorrelated.
    \item \textbf{In-class compactness}: The features of samples from the same class are similar and correlated to each other.
\end{itemize} 
On one hand, for the aim of \emph{cross-class discrimination}, $R(\bar{\boldsymbol{Y}} ,\varepsilon )$ should be as large as possible, so that the whole features could span a space with relatively large possible volume. On the other hand, for the aim of \emph{in-class compactness}, ${R^d}(\bar{\boldsymbol{Y}}|\boldsymbol{\Pi} ,\varepsilon )$ of each class should be as small as possible to span a small volume of space to be maximally coherent. With the above consideration, we need to learn task-relevant semantic features with clear subspace structures (i.e., \emph{cross-class discrimination} and \emph{in-class compactness}), in order to support executing classification task directly on the feature space. To this end, we train the encoding network $e(\cdot,\theta)$ to extract and transmit semantic features from original data $\boldsymbol{S}$ via maximizing the coding rate reduction objective, i.e., 
\begin{align}\label{rate reduction}
\Delta R(\bar{\boldsymbol{Y}}|\boldsymbol{\Pi} ,\varepsilon )= R(\bar{\boldsymbol{Y}} ,\varepsilon ) - {R^d}(\bar{\boldsymbol{Y}}|\boldsymbol{\Pi} ,\varepsilon ).
\end{align}

\begin{remark}
With the loss function in \eqref{rate reduction}, the volume of the space spanned by the whole features $R(\bar{\boldsymbol{Y}} ,\varepsilon )$ is maximized, while the volume of the spaces spanned by each sub-class features ${R^d}(\bar{\boldsymbol{Y}}|\boldsymbol{\Pi} ,\varepsilon )$ is minimized. In such a way, the obtained features from different classes lie in independent subspaces, which could naturally support the goal of cross-class discrimination directly on the feature space. Specifically, it is proofed in \cite{YYu2020} that, considering the obtained features $\bar{\boldsymbol{Y}}_{j_1}$ and $\bar{\boldsymbol{Y}}_{j_2}$ of samples from different classes $j_1$ and $j_2$, we have $(\bar{\boldsymbol{Y}}_{j_1})^{\rm T}\bar{\boldsymbol{Y}}_{j_2}=\mv{0}$, $\forall ~1 \le j_1 \le j_2 \le J$, i.e., $\bar{\boldsymbol{Y}}_{j_1}$ and $\bar{\boldsymbol{Y}}_{j_2}$ are orthogonal. We could also use sphere packing to understand the rate reduction principle. As shown in Figs. \ref{sphere packing}(b) and \ref{sphere packing}(c), $R(\bar{\boldsymbol{Y}} ,\varepsilon )$ in \eqref{rate distortion whole} indicates the total number of balls in the spanned space, while ${R^d}(\bar{\boldsymbol{Y}}|\boldsymbol{\Pi} ,\varepsilon )$ in \eqref{rate distortion class} indicates the number of solid balls in subspaces $S_1$ and $S_2$ (or  $S_1'$ and $S_2'$). In such a way, $\Delta R(\bar{\boldsymbol{Y}}|\boldsymbol{\Pi} ,\varepsilon)$ in \eqref{rate reduction} is the number of hollow balls, which denotes the rate reduction. Maximizing the rate reduction will maximize the number of hollow balls, which leads to the learned features in Fig. \ref{sphere packing}(b) with orthogonal subspaces. In such a case, the desired properties of cross-class discrimination and in-class compactness are achieved. With rate reduction maximization, the obtained features may have clear subspace structures, which depend less on the label information to efficiently support downstream classification tasks as well as defend against label corruption as compared with the features learned via directly fitting labels (based on cross-entropy or information bottleneck principles \cite{YYu2020}). This could be an advantage in the circumstance when the label information is very noisy. In Section V, we will conduct experiments to verify such advantage of defending against label corruption.
\end{remark}

\subsubsection{MSE Minimization}
Next, we further introduce the MSE minimization part of the loss function for data recovery. Specifically, the obtained features $\bar{\boldsymbol{Y}}$ are also expected to be informative enough to capture the semantic details for recovering the source data $\boldsymbol{S}$. To fulfill such goal, we invoke MSE as the second part of the loss function, which is defined as 
\begin{align}\label{MSE}
{\rm{MSE}}(\boldsymbol{S},\hat{ \boldsymbol{S}}) = \frac{1}{NB}{\sum\nolimits_{n = 1}^N {\| {{\boldsymbol{s}_n} - {{\hat{ \boldsymbol{s}}}_n}} \|} ^2}.
\end{align}
By minimizing the MSE between the original data $\boldsymbol{S}$ and the recovered data $\hat{ \boldsymbol{S}}$, the encoder and the decoder are expected to accomplish the pixel-level image reconstruction task efficiently.

Finally, with \eqref{rate reduction} and \eqref{MSE}, the overall loss function for training the deep JSCC framework is presented as
\begin{align}\label{overall loss}
\mathrm{Loss}(\omega, \theta) &=  -\Delta R(\bar{\boldsymbol{Y}}|\boldsymbol{\Pi} ,\varepsilon )+\beta {\rm{MSE}}(\boldsymbol{S},\hat{ \boldsymbol{S}}) \nonumber \\
&=-\Delta R \Big(p \big( {e{(\boldsymbol{S},\theta) \big)}}+\hat{\boldsymbol{n}}|\boldsymbol{\Pi} ,\varepsilon \Big)+\beta{\rm{MSE}}\Big(\boldsymbol{S},d \Big(p \big({e{(\boldsymbol{S},\theta) }}\big) +\hat{\boldsymbol{n}},\omega\Big)\Big),
\end{align}
where $\beta$ is a coefficient to balance the performance of data recovery and classification task execution. It is observed from \eqref{overall loss} that, the training of the parameters  $\theta$ of the encoder will be influenced by both the coding rate reduction maximization and MSE minimization (as both of these parts involve the parameters of the encoder $\theta$), while the training of the parameters $\omega$ of the decoder only relies on the minimization of MSE.

Nevertheless, it is worth noting that the proposed deep JSCC framework in this section is dedicatedly designed for certain deterministic channel condition (specifically, the proposed deep JSCC framework is trained on certain specific SNR) with fixed output feature dimension. To further reduce communication overhead as well as be more robust to variational channel conditions, in the following section, we propose a gated deep JSCC framework training with domain randomization.

\section{Gated Deep JSCC Framework with Domain Randomization}
\begin{figure}[h]
	\centering
	 \epsfxsize=1\linewidth
		\includegraphics[width=16.5cm]{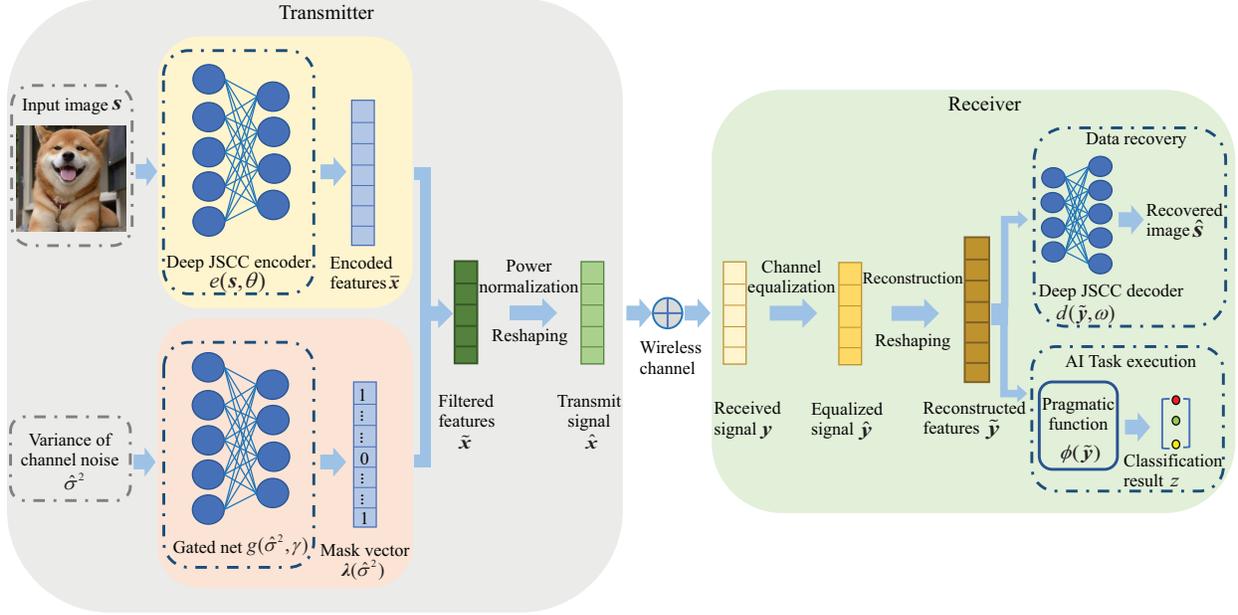}
	\caption{\label{variational model}Illustration of the proposed gated deep JSCC framework for variational channel conditions.}
	\end{figure}

The proposed framework in Fig. \ref{model} is dedicatedly designed for a certain deterministic channel condition with fixed output feature dimension. However, with such a design manner, when the channel condition during inference becomes different from that during training, the performance of the proposed deep JSCC framework will be compromised due to the model mismatch. In such a case, the model needs to be retrained, which suffers from high storage and computation cost in practice. Moreover, it is also interesting to allow the output feature dimension being adaptive to varying channel conditions for saving communication cost. Practically, such consideration is similar to the error control coding in modern communication systems, where a relatively larger code length is needed for correcting the transmission error in worse channel conditions and vice versa. To tackle the above issues, we propose a gated deep JSCC framework training with domain randomization as shown in Fig. \ref{variational model}, where the gated net and domain randomization are invoked for adaptive output activation based on variational channel conditions. In the following, we present the network structure and the training criterion via domain randomization in detail, respectively.

\subsection{Gated Deep JSCC}
To enable the capability of dynamic output feature dimension adaptation under variational channel conditions, we design a gated net for adaptive feature pruning \cite{ZChen2019,Shao2022}. Specifically, the gated net $g({\hat{\sigma} ^2},\gamma )$ is designed as a multilayer perceptron (MLP) with $L$ layers parameterized by $\gamma$, which takes the channel condition (the variance $\hat{\sigma}^2$ of $\hat{\boldsymbol{n}}$ in \eqref{y_equalization}, i.e., $\hat{\sigma}^2={\sigma^2}/{|h|^2}$) as the input, and outputs a binary-valued vector $\boldsymbol{\lambda}(\hat{\sigma}^2) \in \mathbb{R}^{2b}$ with the same dimension as the output features of the encoding net. The gated net is expected to help to activate more output feature dimensions under bad channel conditions to defend against transmission error, otherwise less dimensions will be activated for relieving the communication overhead.

\begin{remark}\label{gatednet}
According to the properties of MLP \cite{IGoodfellow2016,Shao2022}, $g({\hat{\sigma}^2},\gamma )$ is a non-negative increasing function with respect to (w.r.t.) the input  $\hat{\sigma}^2$  if we choose activation
functions with non-negative derivative (such as ReLu, LReLu, and Tanh), and impose the weights of the MLP to be non-negative.  
\end{remark}

Next, we transform the output of $g({\hat{\sigma}^2},\gamma )$ into a binary-valued mask vector $\boldsymbol{\lambda}(\hat{\sigma}^2)$. Specifically, given a predetermined threshold $\lambda_0$, the $i$-th output dimension of the gated net $g_i({\hat{\sigma}^2},\gamma )$ will be deactivated if $g_i({\hat{\sigma}^2},\gamma ) \le \lambda_0$, and correspondingly the $i$-th dimension of $\boldsymbol{\lambda}(\hat{\sigma}^2)$ will be set as $\boldsymbol{\lambda}_i(\hat{\sigma}^2)=0$. Otherwise, we have $\boldsymbol{\lambda}_i(\hat{\sigma}^2)=1$. According to the property revealed in Remark \ref{gatednet}, more output neurons will be activated under bad channel conditions (i.e., high $\hat{\sigma}^2$) and vice versa.

Finally, we perform element-wise product on the outputs of the gated net $\boldsymbol{\lambda}(\hat{\sigma}^2)$ and the deep JSCC encoder $\bar{\boldsymbol {x}}$ to obtain the final activated features. Before proceeding, we first normalize each row of the weight matrix $\boldsymbol{W}^{\rm e}$ of the last layer in the encoder DNN $e(\cdot,\theta )$, then the final filtered output ${\tilde {\boldsymbol{x}}}$ of the gated deep JSCC encoder is shown as
\begin{align}
\tilde {\boldsymbol{x}} = {\boldsymbol{\lambda} (\hat{\sigma}^2)} \circ \tilde{\boldsymbol{W}^{\rm e}} \tilde e({\boldsymbol{s}},\theta ),
\end{align}
where $\tilde{\boldsymbol{W}^{\rm e}}$ denotes the normalized weight matrix of $\boldsymbol{W}^{\rm e}$, and $\tilde e({\boldsymbol{s}},\theta )$ denotes the output of the penultimate layer (i.e., the $(L-1)$-th layer) in the encoder DNN.

With the designed gated net, the original output of the encoder is adaptively selected based on the output of the gated net. It is worth noting that, once the model is well trained, we deploy the gated net $g({\hat{\sigma}^2},\gamma )$ on both sides of the transmitter and receiver during the inference phase. In such a way, the transmitter only needs to transmit the activated features $\tilde {\boldsymbol{x}}$ according to different channel conditions. After receiving the symbols ${\boldsymbol{y}}$ at the receiver side, the receiver could reconstruct the obtained signal to the same shape as the output of the encoder $\bar{{\boldsymbol{x}}}$ based on the stored gated net $g({\hat{\sigma}^2},\gamma )$. Then the receiver could process the reconstructed features $\tilde {\boldsymbol{y}}$ to perform image recovery and classification tasks.

\subsection{Training via Domain Randomization}
This subsection presents the training criterion of the proposed gated deep JSCC framework. In general, the model trained under a certain specific SNR is only suitable for the channel condition that is  trained on. In order to obtain a general robust model for variational channel conditions, we invoke domain randomization \cite{JTremblay2018} to bridge the gap between the different training and testing environments. Specifically, instead of training our proposed gated deep JSCC framework under a single channel environment (i.e, under a given SNR), we randomly perturb the training environment to a wide range of SNR by uniformly sampling ${\hat{\sigma}^2}$ from the possible range $[{\hat{\sigma}}_{\rm min}^2,{\hat{\sigma}}_{\rm max}^2]$ during the training procedure to make the learned DNNs less sensitive to certain specific SNR. 

\begin{algorithm}[htbp]
	\caption{Algorithm workflow of the proposed gated deep JSCC framework}
	\label{alg2}
	\begin{algorithmic}[1] 
	\Statex *******************************{\it Training Phase}*******************************
	\State \algorithmicrequire A batch of images $\mv{S}$ to be transmitted.
	\State \algorithmicensure The trained gated deep JSCC framework with $e(\cdot,\theta)$, $d(\cdot,\omega)$, and $g(\cdot,\gamma )$.
	\State {\bf Initialization:} Initialize the weights of $e(\cdot,\theta)$, $d(\cdot,\omega)$, and $g(\cdot,\gamma )$ via normal initialization.
	\For{each epoch}
	\State $\!\!\!$Randomly sample $\hat{\sigma}^2$ from $[\hat{\sigma}_{\rm min}^2,\hat{\sigma}_{\rm max}^2]$.
	\Statex $\quad${\bf Transmitter:}
	\State $\quad$ Extract semantic features via $e(\mv{S},\theta)$. 
	\State $\quad$ Discretize $\hat{\sigma}^2$ via \eqref{approximate sigma}, obtain the mask vector $\boldsymbol{\lambda}(\tilde \sigma ^2)$.
	\State $\quad$ Obtain activated features, perform power normalization, and then transmit the symbols over wireless channels.
	\Statex $\quad${\bf Receiver:}
	\State $\quad$ Perform channel equalization, reconstruct the features based on $\boldsymbol{\lambda}(\tilde \sigma ^2)$.
	\State $\quad$ Perform image recovery via $d(\tilde{\boldsymbol{Y}},\omega)$ and image classification via $\phi(\tilde{\mv{Y}})$.
    \State Compute loss function value in \eqref{overall loss}, and train $e(\cdot,\theta)$, $d(\cdot,\omega)$, and $g(\cdot,\gamma )$.
	\EndFor
	\Statex *******************************{\it Inference Phase}*******************************
	\State \algorithmicrequire An image $\mv{s}$ to be transmitted.
	\State \algorithmicensure The recovered image $\hat{\boldsymbol{s}}$ and the classification result ${z}$.
	\Statex {\bf Transmitter:}
	\State $\quad$ Deploy the learned $e(\cdot,\theta)$ and $g(\cdot,\gamma )$. Obtain semantic features $\mv{x}$ and $\boldsymbol{\lambda}(\hat{\sigma} ^2)$.
	\State $\quad$ Obtain activated features, perform power normalization, and then transmit the symbols $\hat{\boldsymbol{x}}$ in \eqref{nomarlized x} over wireless channels.
	\Statex {\bf Receiver:}
	\State $\quad$ Deploy the learned $d(\cdot,\omega)$  and $g(\cdot,\gamma )$. 
	\State $\quad$ Perform channel equalization, and reconstruct the features $\tilde{\boldsymbol{y}}$ based on $\boldsymbol{\lambda}(\hat{\sigma} ^2)$.
	\State $\quad$ Perform image recovery via $d(\tilde{\boldsymbol{y}},\omega)$ and image classification via $\phi(\tilde{\mv{y}})$.
	\end{algorithmic}
\end{algorithm}

It is worth noting that, generally, it is non-trivial for the DNNs to converge efficiently considering a continuous random variable as its input. To tackle such problem, during the training phase, we discretize the possible range of ${\hat{\sigma}^2}$ into $k$ segments, and then approximate the sampled ${\hat{\sigma}^2}$ to its nearby discretization point $\tilde{\sigma}^2$ via 
\begin{align}\label{approximate sigma}
{\tilde \sigma ^2} = \frac{{{\hat{\sigma}}_{\rm max}^2 - {\hat{\sigma}}_{\rm min}^2}}{{2k}} + \frac{{{\hat{\sigma}}_{\rm max}^2 - {\hat{\sigma}}_{\rm min}^2}}{k}\left\lfloor {\frac{{{\hat{\sigma} ^2}k}}{{{\hat{\sigma}}_{\rm max}^2 - {\hat{\sigma}}_{\rm min}^2}}} \right\rfloor.
\end{align}
Finally we feed the discretization point $\tilde{\sigma}^2$ into the gated net  $g({\tilde \sigma ^2},\gamma )$ to obtain the mask vector $\boldsymbol{\lambda}(\tilde \sigma ^2)$.
During the inference phase, we input the real value of ${\hat{\sigma}^2}$ into the gated net to perform adaptive feature activation. The whole algorithm workflow for the proposed gated deep JSCC framework with domain randomization is summarized in Algorithm \ref{alg2}.

\begin{table}[h]
	\caption{\label{DNN structure mnist}Structures of the encoder and decoder DNNs for MNIST}
	\centering
	\begin{tabular}{c|c|c}
	\hline
									  & \textbf{Layer}                         & \textbf{ Output dimensions} \\ \hline
	\multirow{4}{*}{\textbf{Encoder}} & Convolutional layer+LReLu              & 16$\times$16$\times$64                   \\
									  & Convolutional layer+BN+LReLu           & 8$\times$8$\times$128                    \\
									  & Convolutional layer+BN+LReLu           & 4$\times$4$\times$256                    \\
									  & Convolutional layer                    & 1$\times$1$\times$648                    \\ \hline \hline
	\multirow{4}{*}{\textbf{Decoder}} & Transposed convolutional layer+BN+ReLu & 4$\times$4$\times$256                    \\
									  & Transposed convolutional layer+BN+ReLu & 8$\times$8$\times$128                    \\
									  & Transposed convolutional layer+BN+ReLu & 16$\times$16$\times$64                   \\
									  & Transposed convolutional layer+Tanh    & 32$\times$32$\times$1                    \\ \hline
	\end{tabular}
	\vspace{-20pt}
	\end{table}

\section{Numerical Results}
This section presents numerical results to validate the performance of our proposed deep JSCC designs. Specifically, the experiments are conducted on MNIST \cite{Y. LeCun} and CIFAR-10 \cite{AKrizhevsky2009} datasets with NVIDIA RTX 3090 graphics processing units (GPUs). In the following, we introduce the experimental setups and simulation results, respectively.

\subsection{Experimental Setups}
\subsubsection{Datasets}
We first briefly introduce the MNIST and CIFAR-10 datasets as follows.
\begin{itemize}
\item {\bf MNIST}: MNIST  dataset \cite{Y. LeCun} collects images of handwritten digits from ``0'' to ``9'' each with $28\time28$ grey pixels. It has a training set of 60,000 samples, and a test set of 10,000 samples. 
\item {\bf CIFAR-10}: CIFAR-10  dataset \cite{AKrizhevsky2009}  consists of colorful images with $32\times32\times3$ pixels of diverse real-world objects from 10 different classes (such as bird, cat). It has 50000 images for training and 10000 images for testing.
\end{itemize}

\begin{table}[h]
	\caption{\label{DNN structure cifar10}Structures of the encoder and decoder DNNs for CIFAR-10}
	\centering
	\begin{tabular}{c|c|c}
	\hline
									  & \textbf{Layer}                         & \textbf{ Output dimensions} \\ \hline
	\multirow{4}{*}{\textbf{Encoder}} & ResBlock down               & 16$\times$16$\times$64                   \\
									  & ResBlock down          & 8$\times$8$\times$128                    \\
									  & Resblock down           & 4$\times$4$\times$256                    \\
									  & Convolutional layer                    & 1$\times$1$\times$700                    \\ \hline \hline
	\multirow{4}{*}{\textbf{Decoder}} & Transposed convolutional layer+BN+ReLu & 4$\times$4$\times$256                    \\
									  & Transposed convolutional layer+BN+ReLu & 8$\times$8$\times$128                    \\
									  & Transposed convolutional layer+BN+ReLu & 16$\times$16$\times$64                   \\
									  & Transposed convolutional layer+Tanh    & 32$\times$32$\times$3                    \\ \hline
	\end{tabular}
\end{table}

\subsubsection{DNN Structures and Hyper-parameter Settings}
The structures of the encoder and decoder DNNs for MNIST and CIFAR-10 datasets are summarized in Tables \ref{DNN structure mnist} and \ref{DNN structure cifar10}, respectively, where BN denotes the batch normalization layer, LReLu denotes the LeakyReLu activation function, ResBlock down is the same as down-sampling ResBlock in the ResNet \cite{KHe2016}. It is worth noting that we first resize the images in MNIST as $32\times32$ before feeding them into the DNNs. Furthermore, we set the key hyper-parameters as follows. We set the coding distortion as $\epsilon^2=0.5$, the learning rate as $1.5{\rm e}^{-4}$, the batch size as $2048$ for MNIST and $1600$ for CIFAR-10, and the negative slope of LReLu as $0.2$. Moreover, we set $b/B=0.316$ and $0.114$ for MNIST and CIFAR-10 datasets, respectively, and thus the output dimensions of the encoders are $648$ and $700$, respectively. Also, we set the output dimensions of the decoders for MNIST and CIFAR-10 as $1$ and $3$ according to the number of their color channels. 

\begin{figure}[h]
	\centering  
	\subfigure[Raw data]{
\includegraphics[width=0.42\linewidth]{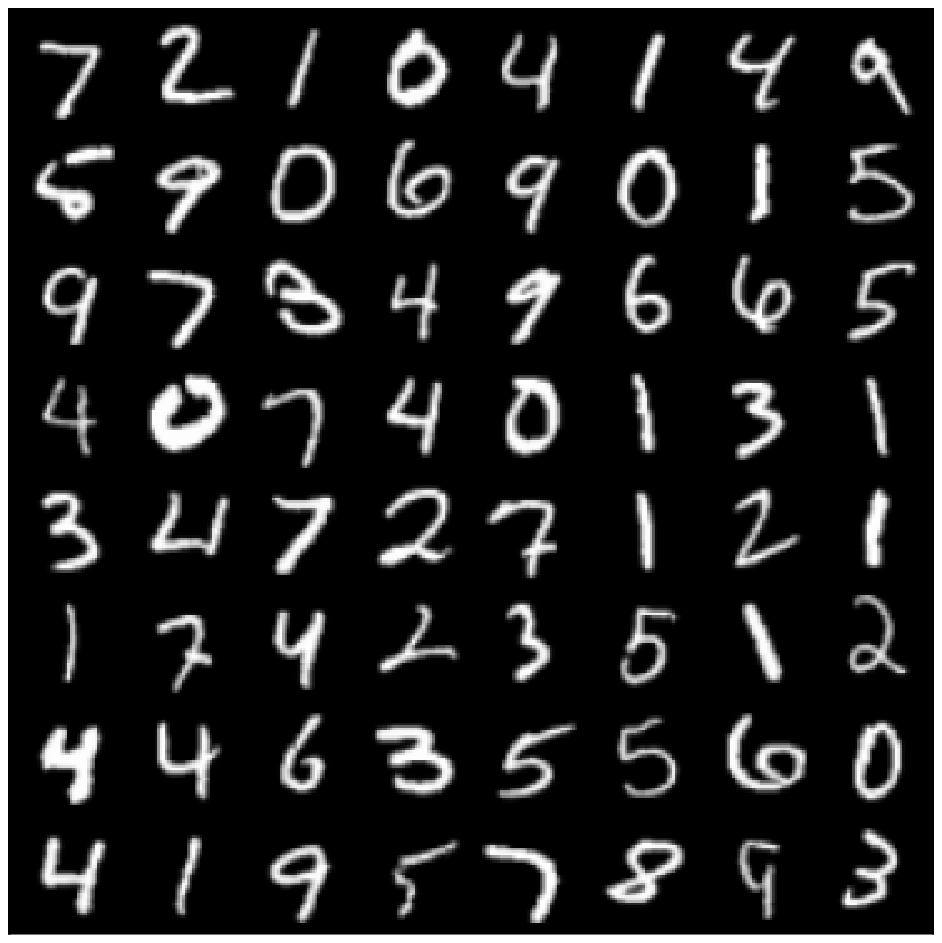}
\label{Mnist Raw data}
}
	\subfigure[Transmitting through AWGN channels]{
\includegraphics[width=0.42\linewidth]{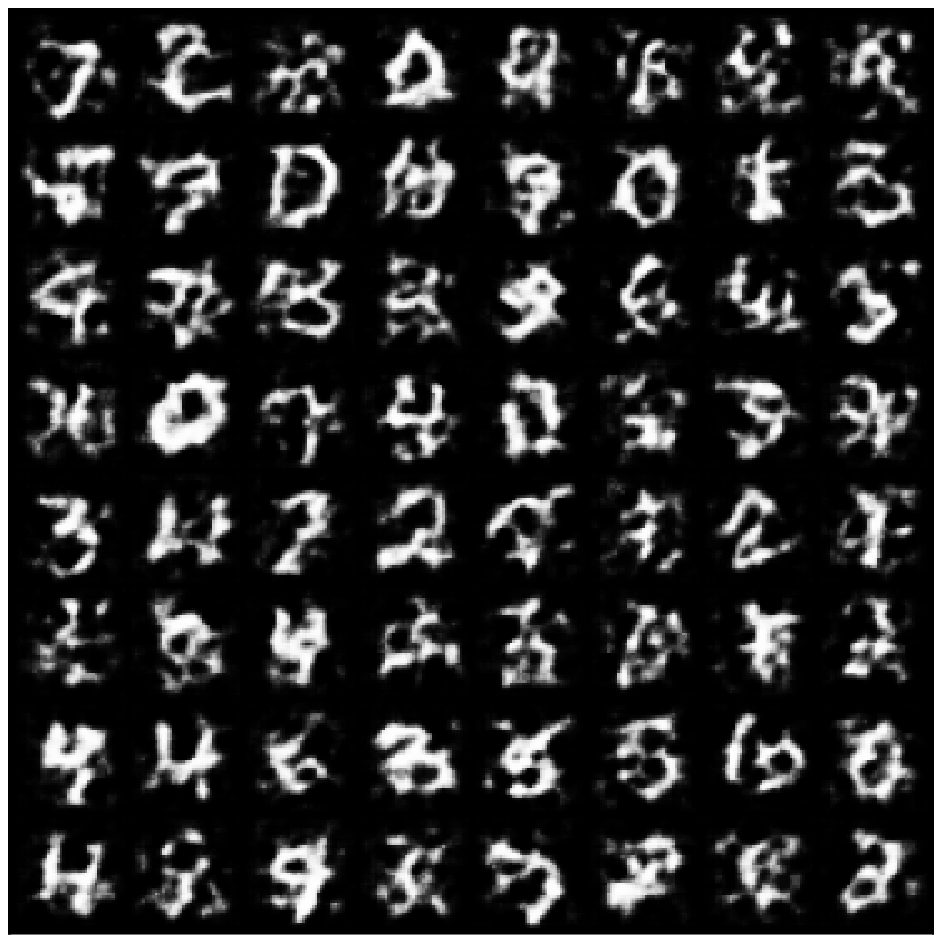}
\label{Mnist AWGN wo channel}
}
	  \\
	\subfigure[Transmitting through Rayleigh fading channels]{
\includegraphics[width=0.42\linewidth]{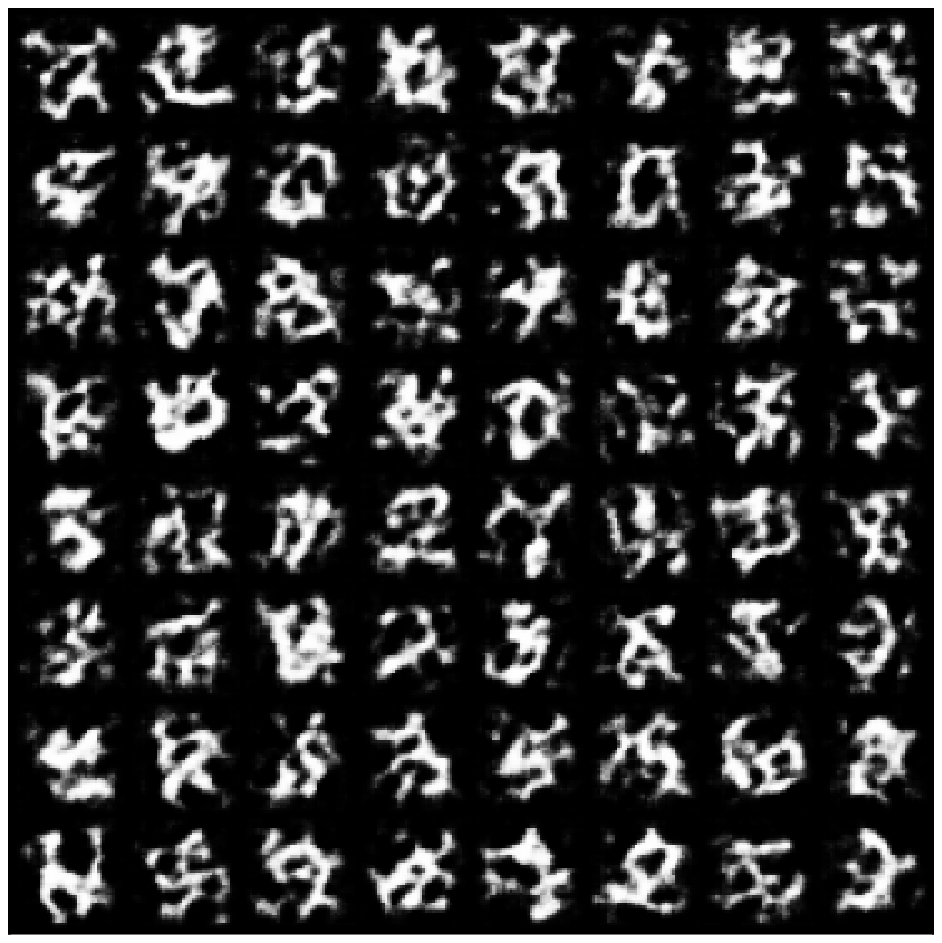}
\label{Mnist Rayleigh wo channel}
}
	\subfigure[Transmitting through Rician fading channels]{
\includegraphics[width=0.42\linewidth]{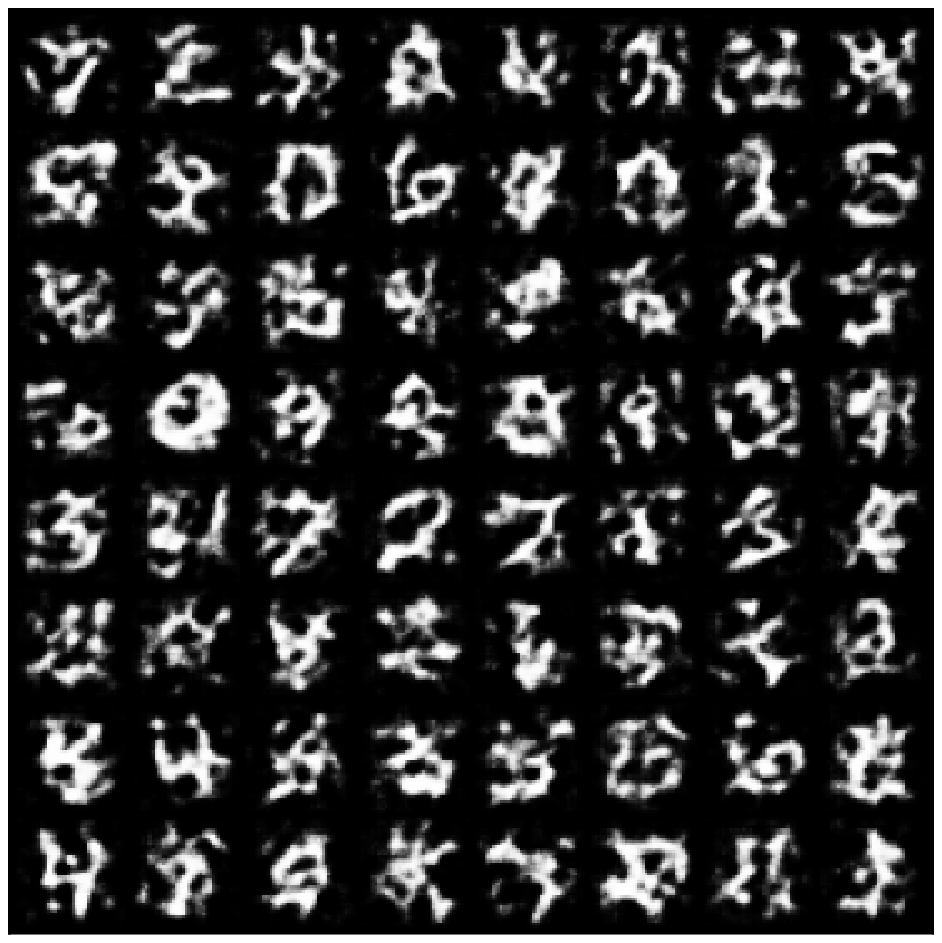}
\label{Mnist Rician wo channel}
}
	\caption{Effects of channel impairments on MNIST image transmission. }
	\label{raw data mnist}
\end{figure}

\begin{figure}[h]
	\centering  
	\subfigure[Raw data]{
\includegraphics[width=0.42\linewidth]{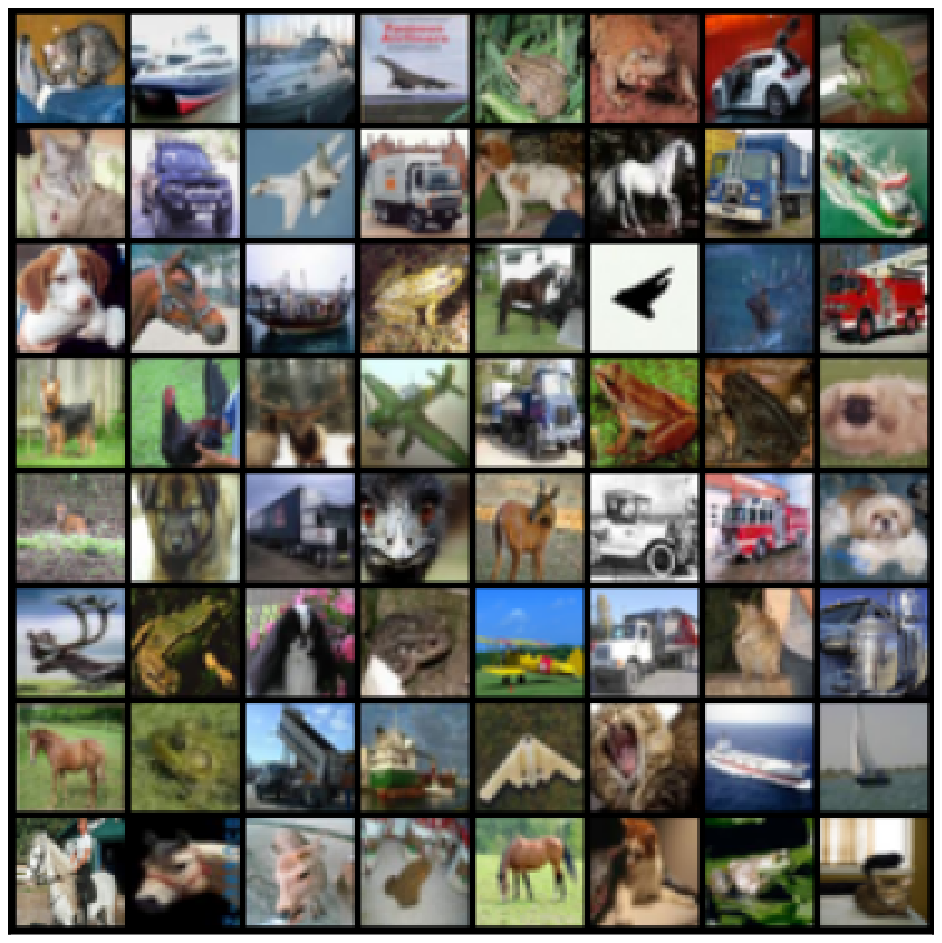}
\label{Cifar10 raw data}
}
	\subfigure[Transmitting through AWGN channels]{
\includegraphics[width=0.42\linewidth]{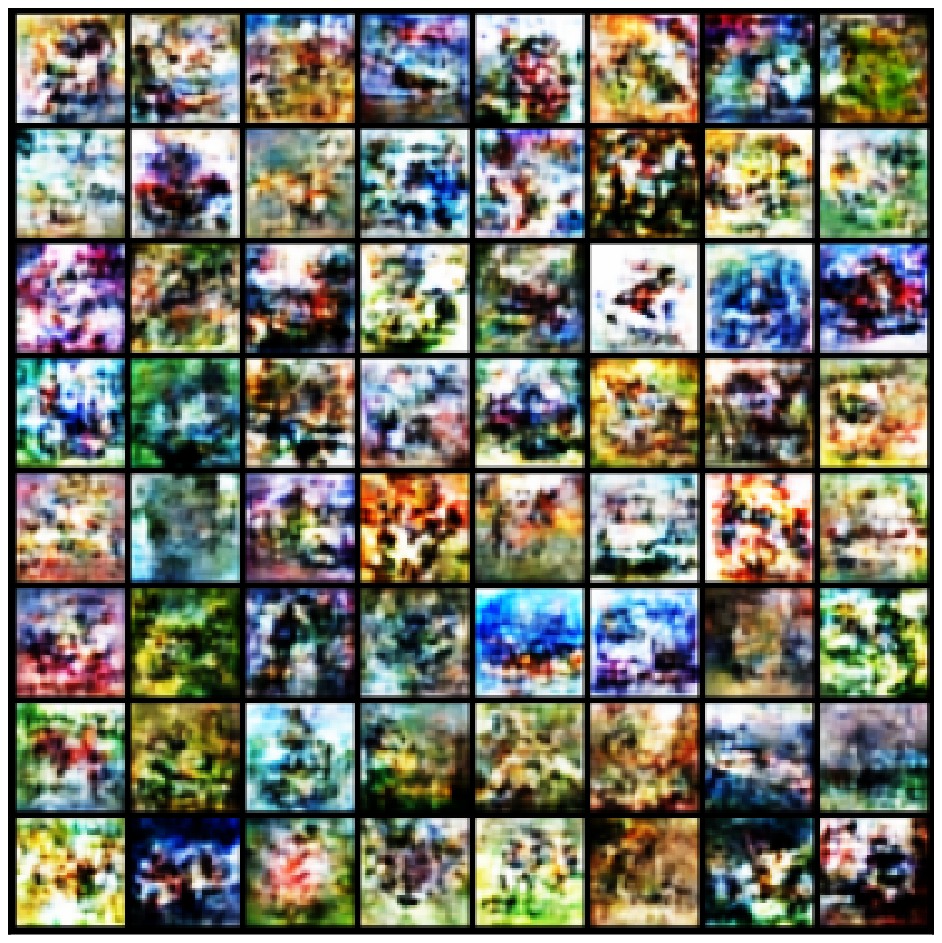}
\label{Cifar10 AWGN wo channel}
}
	  \\
	\subfigure[Transmitting through Rayleigh fading channels]{
\includegraphics[width=0.42\linewidth]{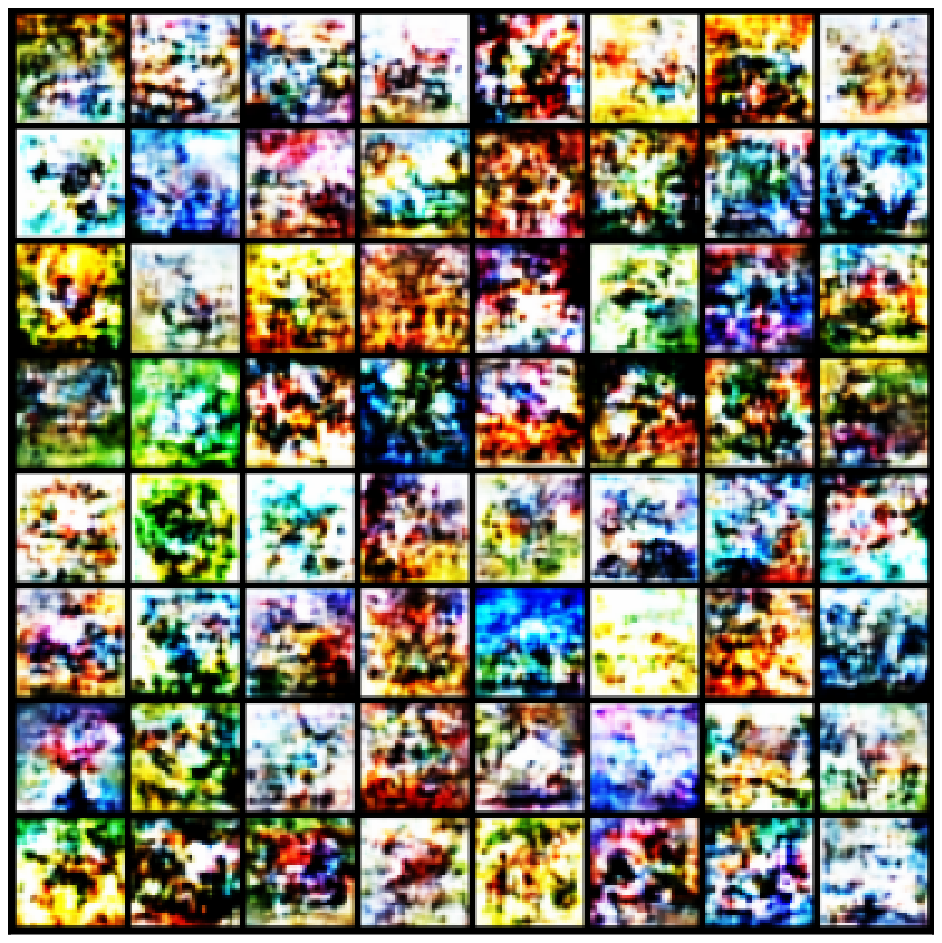}
\label{Cifar10 Raylegh wo channel}
}
	\subfigure[Transmitting through Rician fading channels]{
\includegraphics[width=0.42\linewidth]{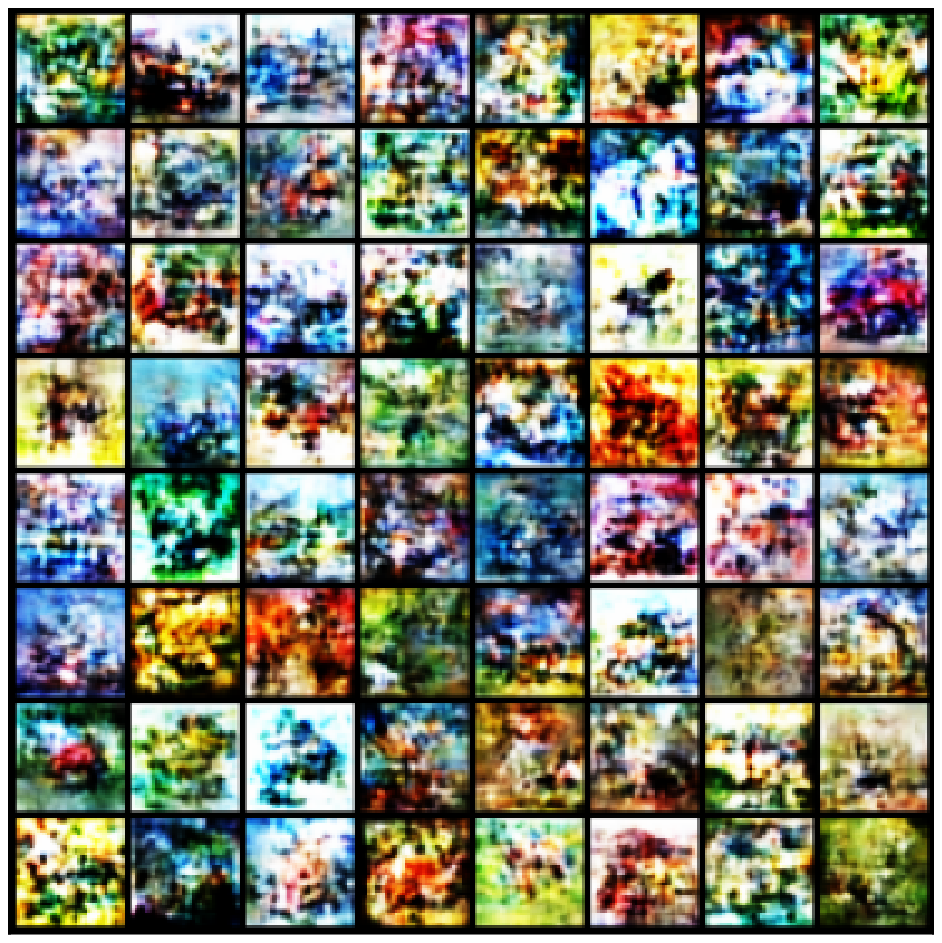}
\label{Cifar10 Rician wo channel}
}
	\caption{Effects of channel impairments on CIFAR-10 image transmission. }
	\label{raw data cifar10}
\end{figure}

\subsubsection{Performance Metrics}
We evaluate the performance of image recovery task by PSNR as defined in \eqref{PSNR}, and image classification task by classification accuracy. Particularly, to exemplify  $\phi(\cdot)$ for classification, we adopt nearest subspace classifier \cite{YMa2007,YYu2020} to perform the classification task directly on the feature space, as the learned features have very clear subspace structures. Specifically, for each class of obtained features $\bar{\boldsymbol{Y}}_j$, let $\boldsymbol{\mu}_j$ and $\boldsymbol{V}_j$ represent the vectors of mean values and the first $p_j$ principle components of $\bar{\boldsymbol{Y}}_j$, respectively. Then the predicted label $j'$ of a test image ${\boldsymbol{s}}^{\prime}$ based on its  features $\bar{\mv{y}}'$ is obtained as
\begin{align}
j' = \mathop {\arg \min }\limits_{j \in \{ 1, \ldots J\} } \left\| {(\boldsymbol{I} - {\boldsymbol{V}_j}\boldsymbol{V}_j^H)(\bar{\mv{y}}'- {{\boldsymbol{\mu}} _j})}  \right\|{^2}.
\end{align}

\begin{figure}[h]
	\centering
	\subfigure[Transmitting through AWGN channels]{
	\centering
	\includegraphics[width=0.32\linewidth]{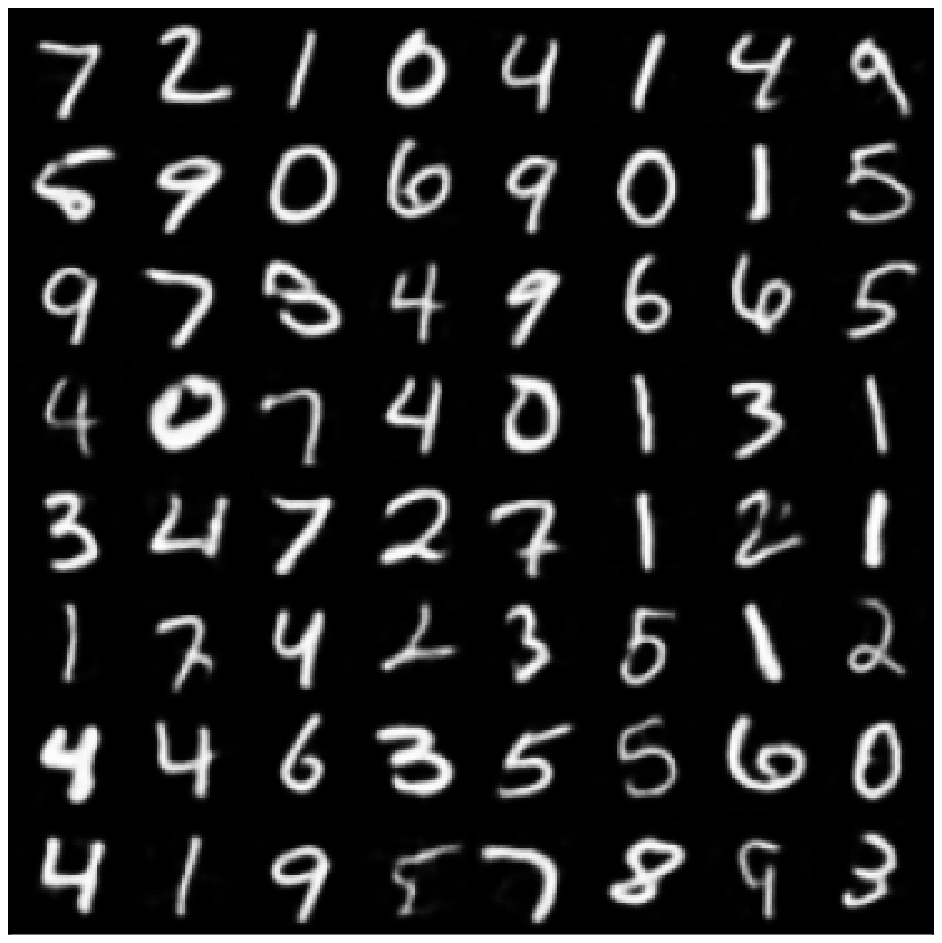}
	\label{mnist awgn w channel}
	}%
	\subfigure[Transmitting through Rayleigh fading channels]{
	\centering
	\includegraphics[width=0.32\linewidth]{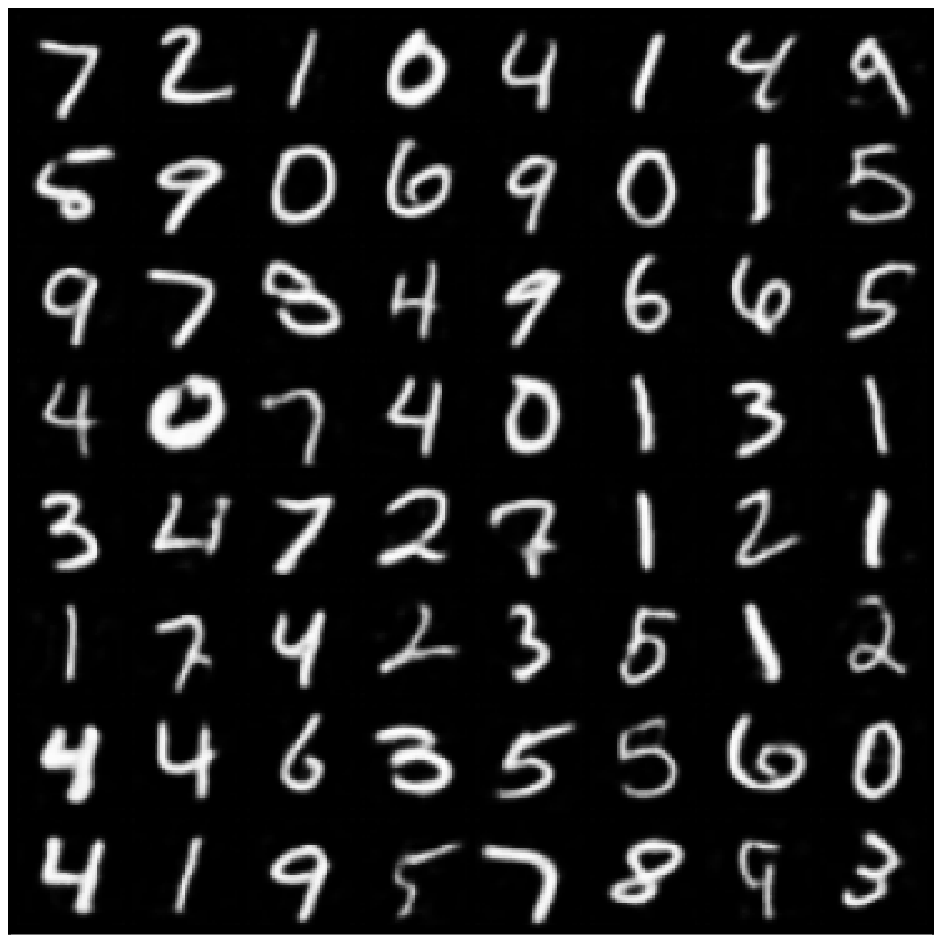}
	\label{mnist rayleigh w channel}
	}%
	\subfigure[Transmitting through Rician fading channels]{
	\centering
	\includegraphics[width=0.32\linewidth]{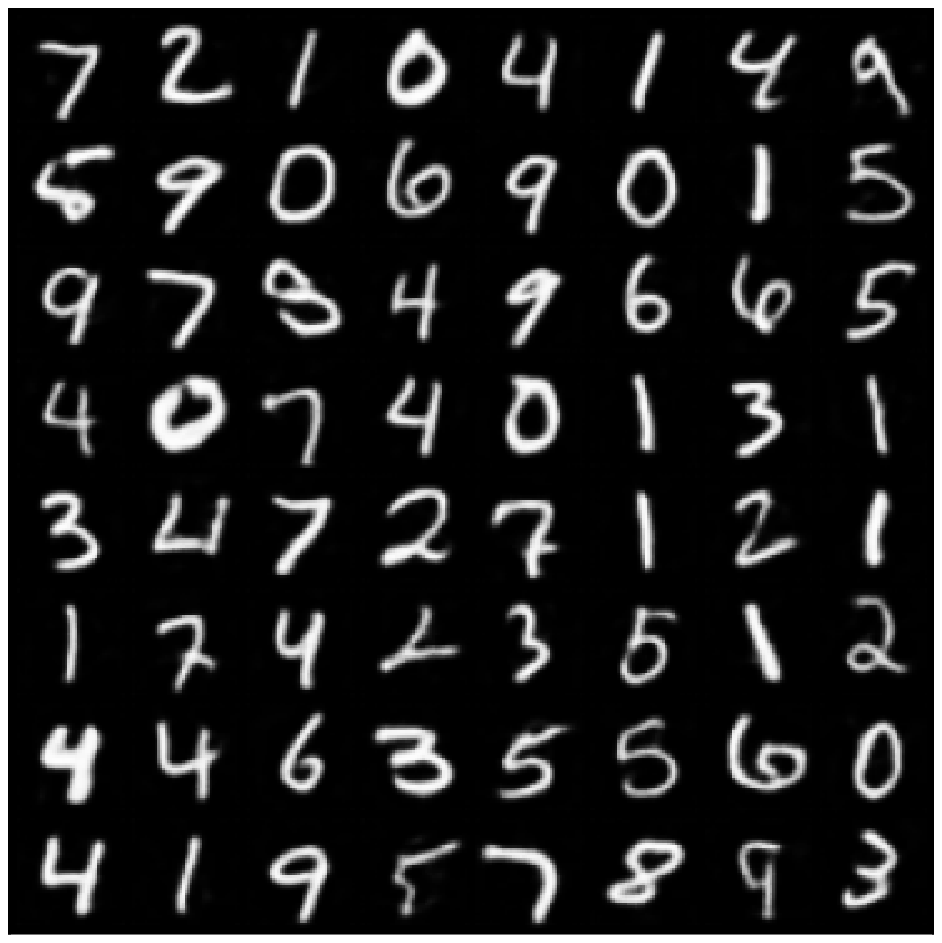}
	\label{mnist rician w channel}
	}%
	\centering
	\vspace{-10pt}
	\caption{Transmitting MNIST images through AWGN, Rayleigh, and Rician fading channels under $0~{\rm dB}$ with the proposed  deep JSCC framework.}
	\label{train on channels mnist}
	\end{figure}

\subsection{Performance Evaluation of the Proposed Deep JSCC Framework}

In this subsection, we present simulation results for validating our proposed deep JSCC framework on performing image recovery and classification task simultaneously. We show the performance of our proposed deep JSCC framework as compared to the following benchmark schemes.
\begin{itemize}
	\item {\bf Deep JSCC with MSE}: We consider minimizing  MSE loss as defined in \eqref{MSE} to train the whole deep JSCC framework in an end-to-end manner.
	\item {\bf Deep JSCC with SSIM}: We consider utilizing SSIM loss \cite{ZWang2004} to train the whole deep JSCC framework, which is defined as
	\begin{align}
		\mathrm{Loss}_{\rm SSIM}(\omega,\theta)&=1-\frac{1}{N}\sum\limits_{n = 1}^N {\rm SSIM}(\boldsymbol{s}_n,\hat{\boldsymbol{s}}_n) \nonumber \\
		&=1-\frac{1}{N}\sum\limits_{n = 1}^N {\rm SSIM}\bigg( \boldsymbol{s}_n,d \Big(p \big({e{(\boldsymbol{s}_n,\theta) }}\big) +\hat{\boldsymbol{n}},\omega\Big) \bigg),
	\end{align}
	where ${\rm SSIM}(\cdot)$ is the function computing structure similarity between two images as
	\begin{align}
		{\rm SSIM}(\boldsymbol{s}_n,\hat{\boldsymbol{s}}_n)=\frac{(2{\rho}_{\boldsymbol{s}_n} {\rho}_{\hat{\boldsymbol{s}}_n}+o_1)(2{\delta}_{\boldsymbol{s}_n \hat{\boldsymbol{s}}_n}+o_2)}{({\rho}^2_{\boldsymbol{s}_n} +{\rho}^2_{\hat{\boldsymbol{s}}_n}+o_1)({\delta}^2_{\boldsymbol{s}_n} +{\delta}^2_{\hat{\boldsymbol{s}}_n}+o_2)},
	\end{align}
	where $\rho_{\boldsymbol{s}_n}$ (or ${\rho}_{\hat{\boldsymbol{s}}_n}$) and ${\delta}_{\boldsymbol{s}_n}$ (or ${\delta}_{\hat{\boldsymbol{s}}_n}$) denote the mean and variance of the pixel values of image ${\boldsymbol{s}_n}$ (or ${\hat{\boldsymbol{s}}_n}$), respectively, ${\delta}_{\boldsymbol{s}_n \hat{\boldsymbol{s}}_n}$ denotes the covariance of two images, and $o_1$ and $o_2$ are two constants.

	\item {\bf SSCC (with JPEG2000 and capacity-achieving channel coding)}: We first consider utilizing JPEG2000 as source coding method. Then we consider capacity-achieving channel coding. Specifically, the source image $\boldsymbol{s}$
	is compressed via JPEG2000 at a rate being equal to the channel capacity $\Omega={\rm log}_2(1+SNR)$. Here, the capacity $\Omega$  is the maximum number of bits per symbol for reliable communication over the discrete memoryless noisy channel. With capacity $\Omega$, we could compress and transmit the source at the maximum rate as
	\begin{align}\label{max rate}
	R_{\rm max}=\frac{b}{B}\Omega.
	\end{align}
	
	Also, there exists a lower bound for transmission rate $R_{\rm min}$, below which the image will be totally distorted to unsuccessful recovery. This is due to the fact that one cannot compress images to an arbitrary low rate for source compression methods such as JPEG2000. If the SNR is low and/or the compression ratio $b/B$ is small, then $R_{\rm min}>R_{\rm max}$ may happen, then the transmitted image cannot be recovered. In such a case, the decoder randomly generates the recovered images. Otherwise, if $R_{\rm min}<R_{\rm max}$, we chose the largest achievable rate of JPEG2000 (which is smaller than $R_{\rm max}$) to compress the images assuming error-free transmission of the compressed bitstreams. It is worth noting that to perform classification via SSCC, we input the obtained compressed image into the deep JSCC encoder and nearest subspace classifier to obtain the classification results similar as in \cite{HXie2022}.
\end{itemize}

\begin{figure}[t]
	\centering
	\subfigure[Transmitting through AWGN channels]{
	\centering
	\includegraphics[width=0.32\linewidth]{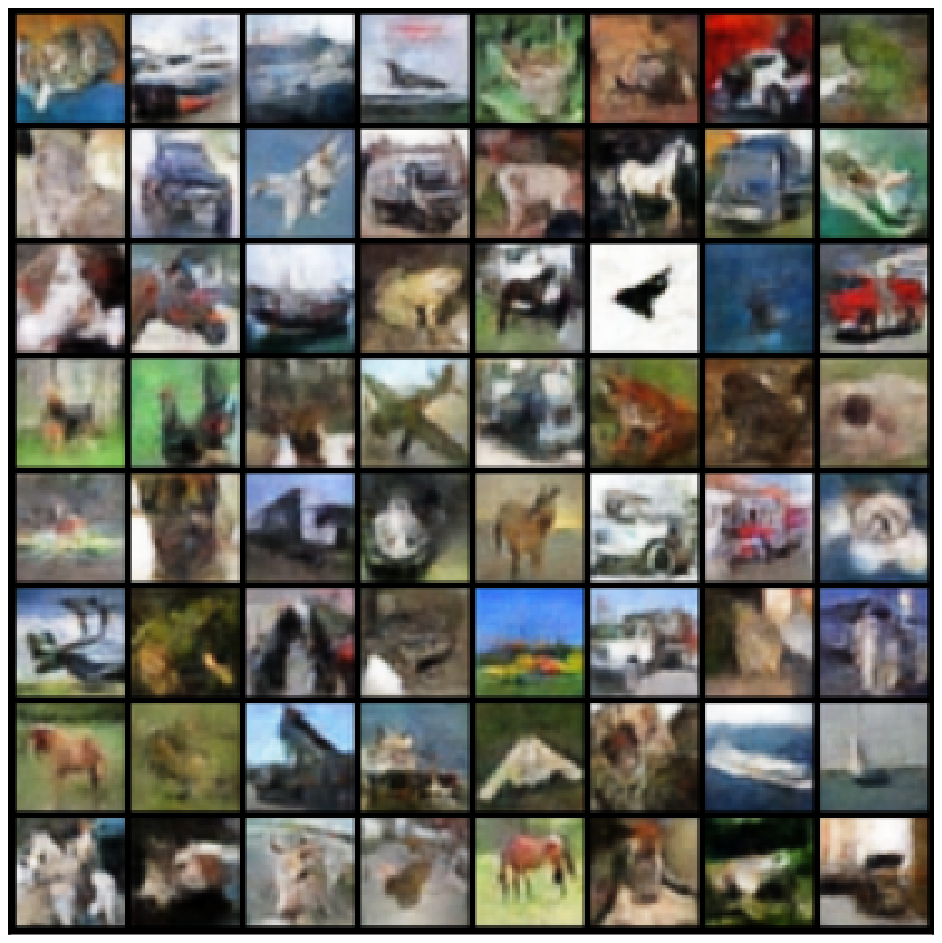}
	\label{cifar10 awgn w channel}
	}%
	\subfigure[Transmitting through Rayleigh fading channels]{
	\centering
	\includegraphics[width=0.32\linewidth]{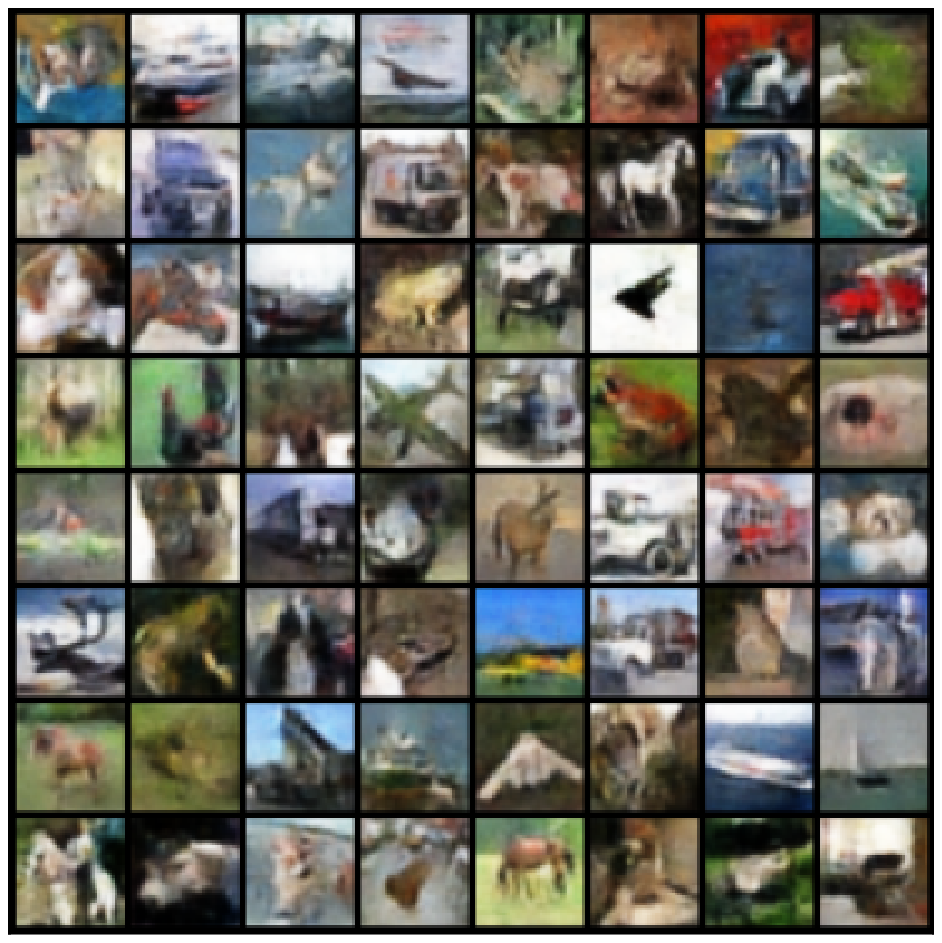}
	\label{cifar10 rayleigh w channel}
	}%
	\subfigure[Transmitting through Rician fading channels]{
	\centering
	\includegraphics[width=0.32\linewidth]{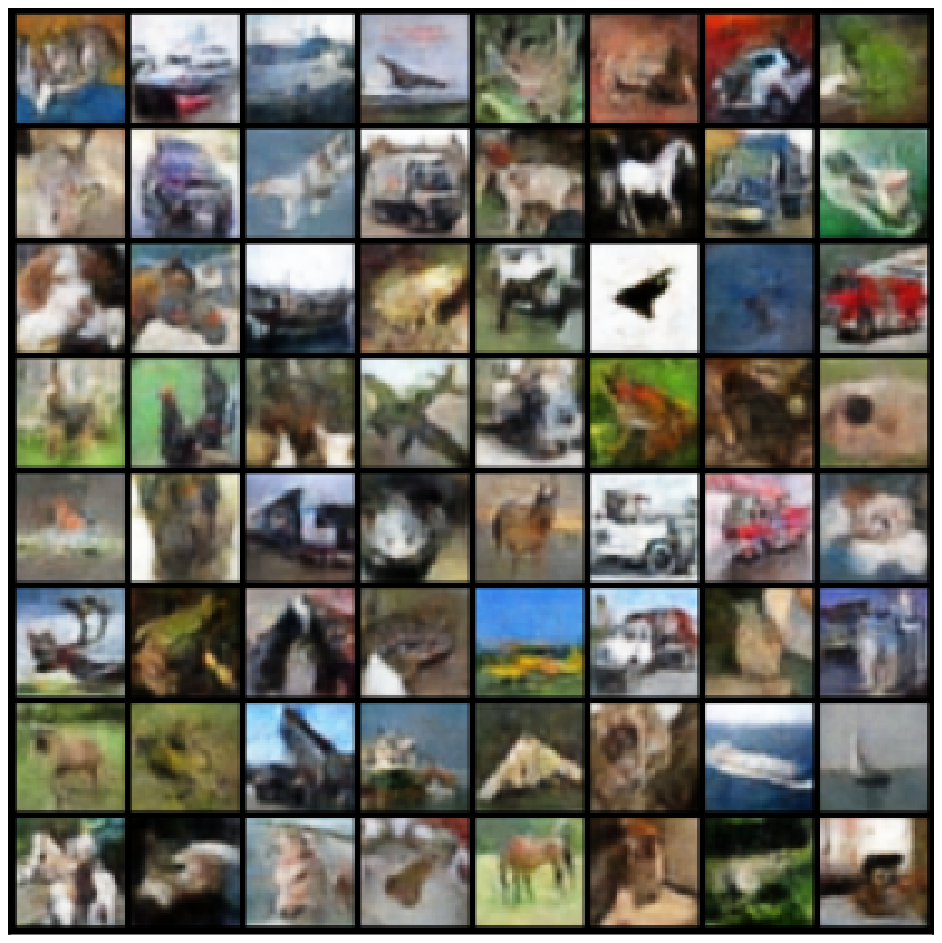}
	\label{cifar10 rician w channel}
	}%
	\centering
	\vspace{-10pt}
	\caption{Transmitting CIFAR-10 images through AWGN, Rayleigh, and Rician fading channels under $0~{\rm dB}$ with the proposed deep JSCC framework.}
	\label{train on channels cifar10}
	\vspace{-10pt}
	\end{figure}

First, we show the effects of channel impairments on wireless image transmission under $0~{\rm dB}$ in Figs. \ref{raw data mnist} and \ref{raw data cifar10}. Specifically, we first train the deep JSCC framework without integrating wireless channels in the whole network structure, then we test the obtained encoder and decoder under wireless transmission (i.e., transmitting the extracted features of the obtained encoder under block fading channels, with AWGN, Rayleigh fading, or Rician fading in each slot, respectively). It is shown in Figs. \ref{raw data mnist} and \ref{raw data cifar10} that channel impairments (i.e., shadowing and fading) cause image distortion in wireless image transmission, and thus compromise the quality of image recovery significantly.

Next, in Figs. \ref{train on channels mnist} and \ref{train on channels cifar10}, we present the results of image recovery visually by the proposed deep JSCC framework under $0~{\rm dB}$, which integrates wireless channels to enable end-to-end training. By comparing Figs. \ref{Mnist AWGN wo channel} and \ref{mnist awgn w channel}, Figs. \ref{Mnist Rayleigh wo channel} and \ref{mnist rayleigh w channel}, and Figs. \ref{Mnist Rician wo channel} and \ref{mnist rician w channel} for MNIST (or comparing Figs. \ref{Cifar10 AWGN wo channel} and \ref{cifar10 awgn w channel}, Figs. \ref{Cifar10 Raylegh wo channel} and \ref{cifar10 rayleigh w channel}, and Figs. \ref{Cifar10 Rician wo channel} and \ref{cifar10 rician w channel} for CIFAR-10), it is shown that the image recovery quality is enhanced significantly by integrating wireless channels into the whole deep JSCC framework to enable end-to-end training. This suggests that the proposed deep JSCC framework is able to properly adapt the encoder and  decoder for semantic feature extraction and transmission to defend channel impairments accordingly.

\begin{figure}[h]
	\centering
	\subfigure[Deep JSCC with MSE]{
	\centering
	\includegraphics[width=0.32\linewidth]{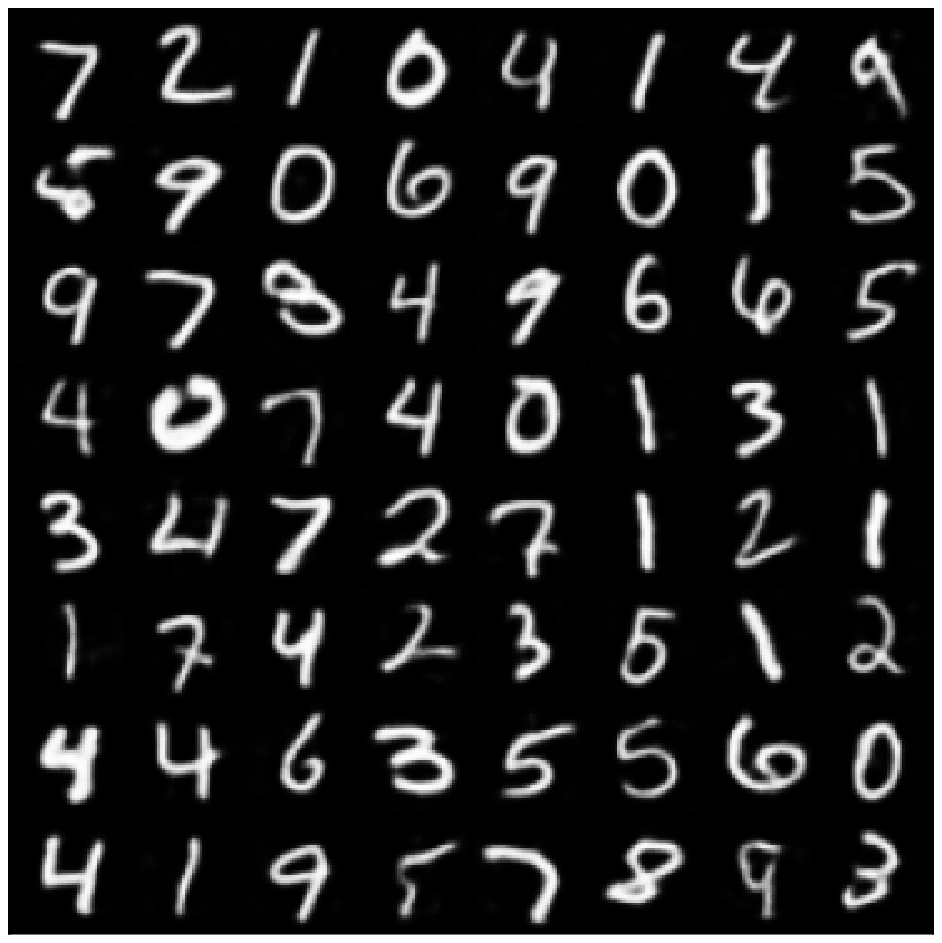}
	\label{mse recovery images for mnist}
	}%
	\subfigure[Deep JSCC with SSIM]{
	\centering
	\includegraphics[width=0.32\linewidth]{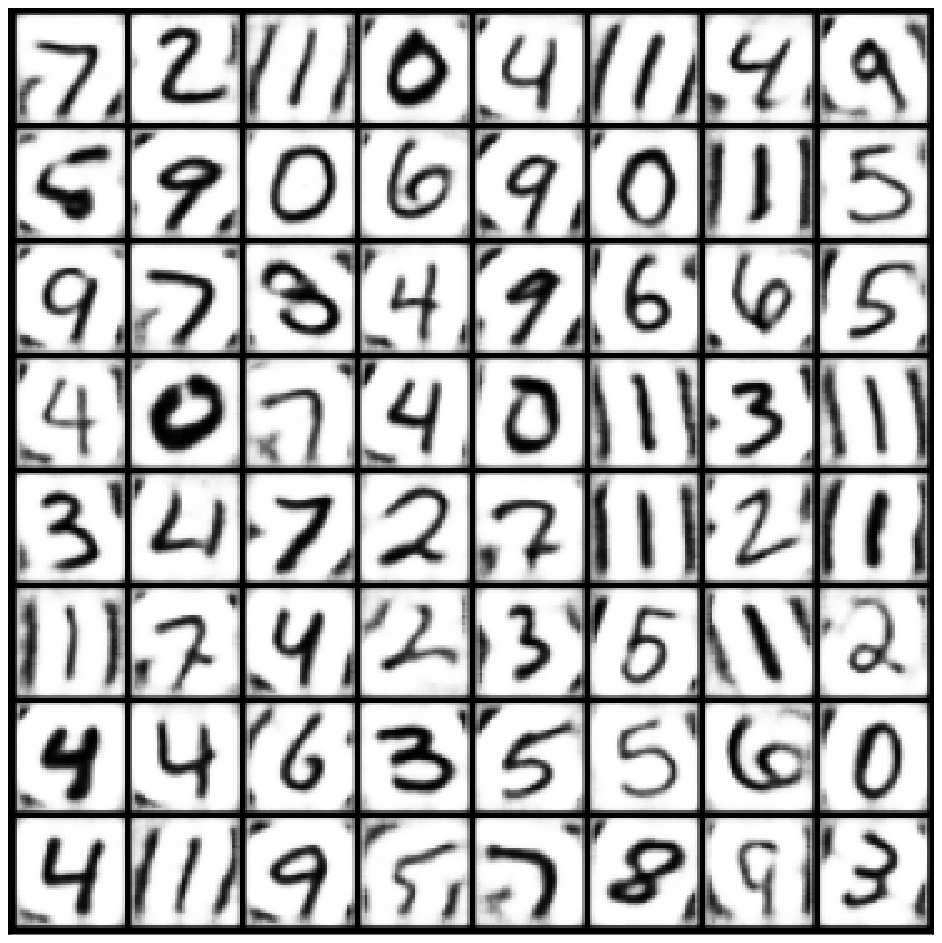}
	\label{ssim recovery images for mnist}
	}%
	\subfigure[SSCC]{
	\centering
	\includegraphics[width=0.32\linewidth]{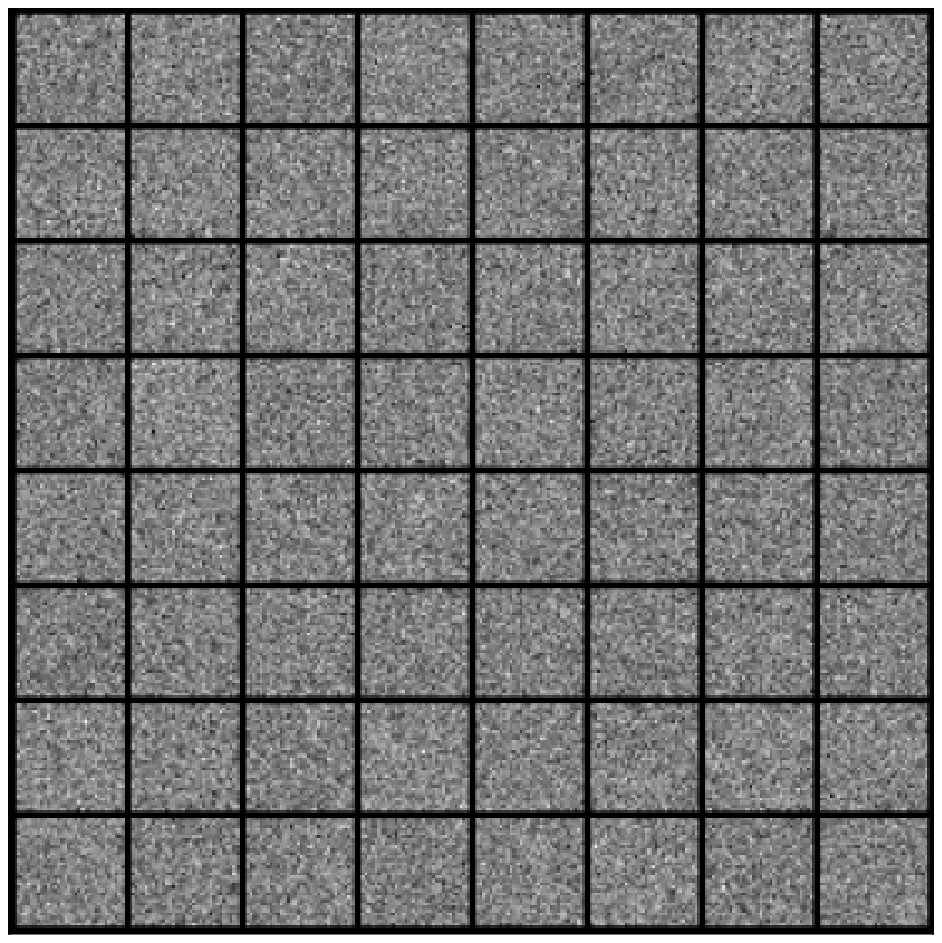}
	\label{sscc recovery images for mnist}
	}%
	\centering
	\vspace{-10pt}
	\caption{Transmitting MNIST images through AWGN  channels under $0~{\rm dB}$ with benchmark schemes.}
	\label{benchmarks recovery images for mnist}
	\vspace{-10pt}
\end{figure}

\begin{figure}
	\centering
	\subfigure[Deep JSCC with MSE]{
	\centering
	\includegraphics[width=0.32\linewidth]{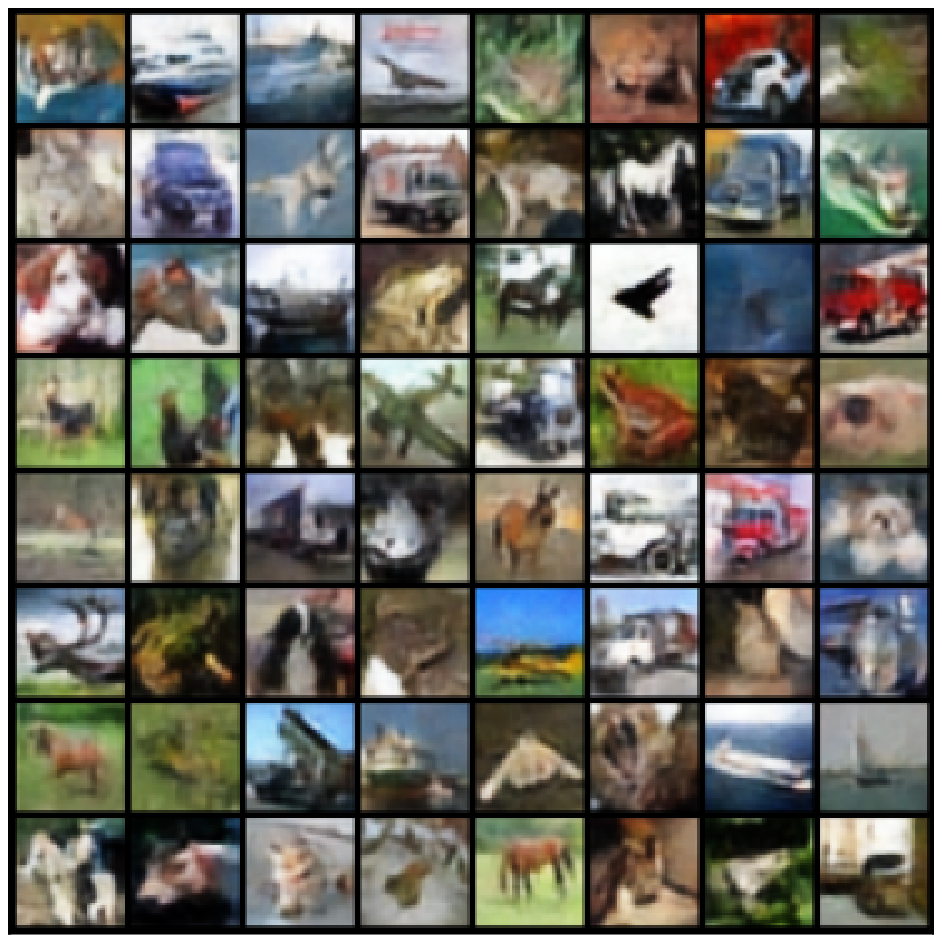}
	\label{mse recovery images for cifar10}
	}%
	\subfigure[Deep JSCC with SSIM]{
	\centering
	\includegraphics[width=0.32\linewidth]{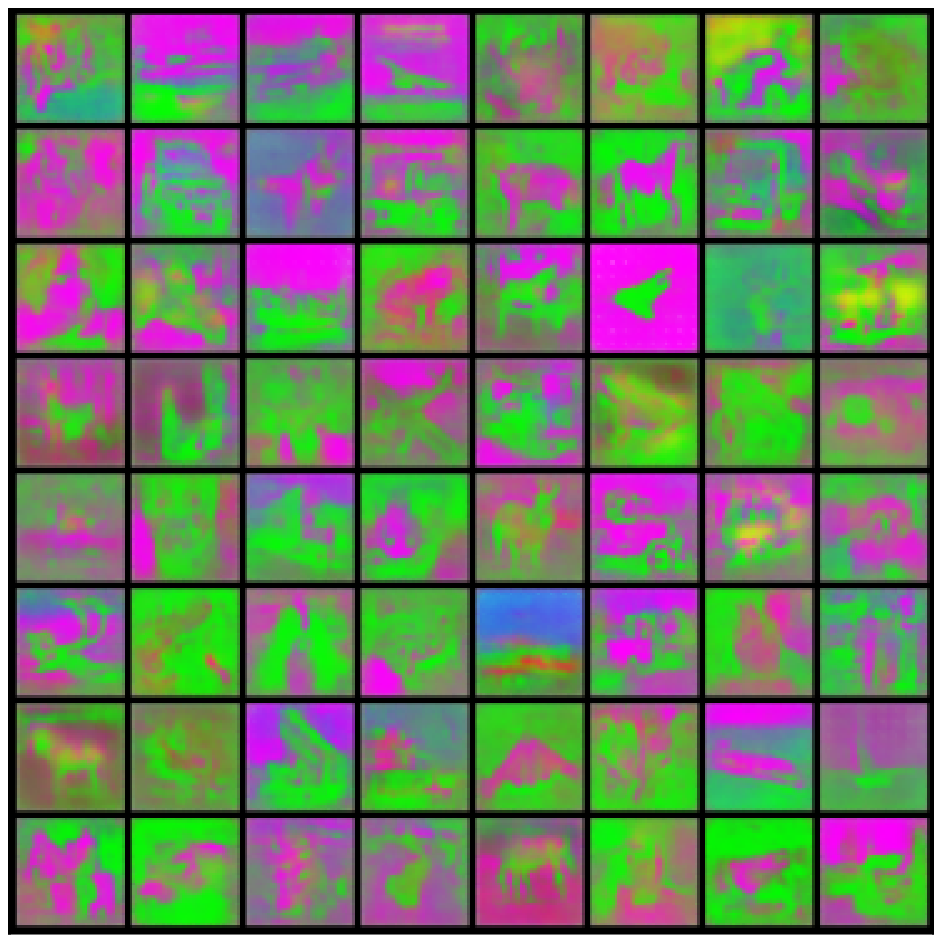}
	\label{ssim recovery images for cifar10}
	}%
	\subfigure[SSCC]{
	\centering
	\includegraphics[width=0.32\linewidth]{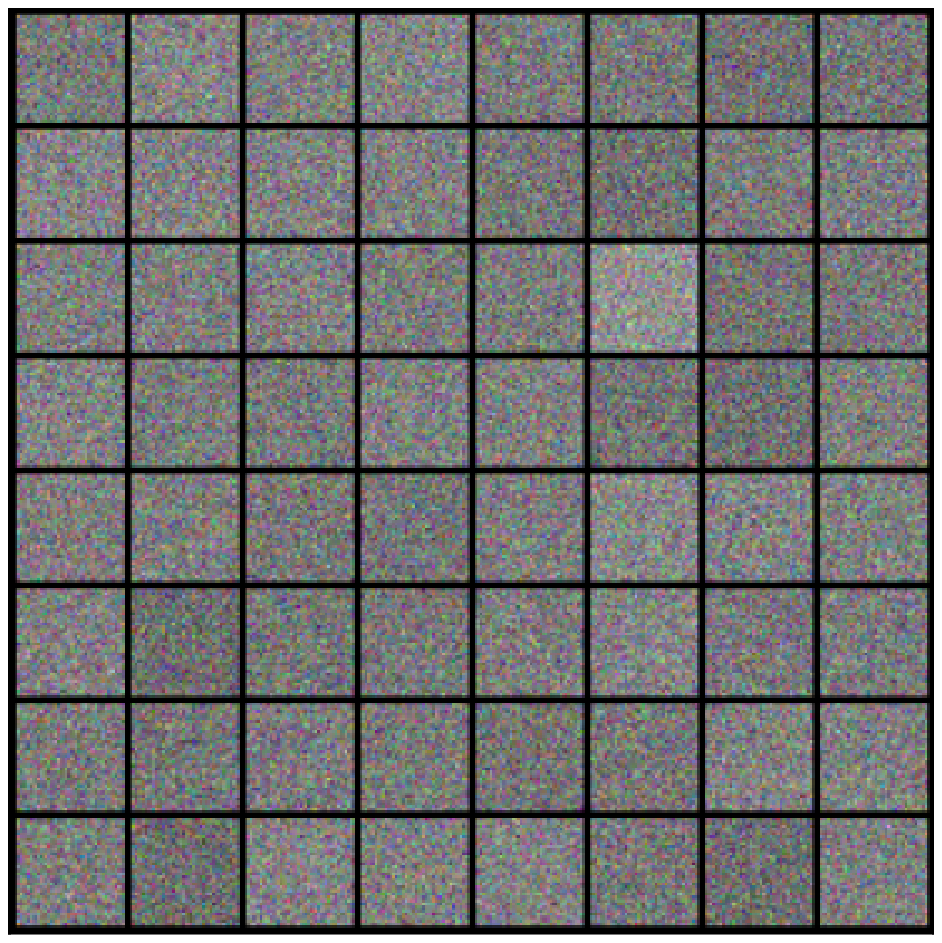}
	\label{sscc recovery images for cifar10}
	}%
	\centering
	\caption{Transmitting CIFAR-10 images through AWGN  channels under $0~{\rm dB}$ with benchmark schemes.}
	\label{benchmarks recovery images for cifar10}
	\vspace{-20pt}
\end{figure}

Then, in Figs. \ref{benchmarks recovery images for mnist} and \ref{benchmarks recovery images for cifar10}, we show the visual results of image recovery on benchmark schemes under AWGN channel with $0~{\rm dB}$. First, by comparing Figs. \ref{mnist awgn w channel} and \ref{mse recovery images for mnist} for MNIST (or Figs. \ref{cifar10 awgn w channel} and \ref{mse recovery images for cifar10} for CIFAR-10), it is observed that the quality of the recovered images of our proposed deep JSCC framework is close to that of the performance upper bound achieved by the deep JSCC with MSE. Second, by comparing Figs. \ref{mnist awgn w channel}, \ref{ssim recovery images for mnist}, and \ref{sscc recovery images for mnist} for MNIST (or Figs. \ref{cifar10 awgn w channel}, \ref{ssim recovery images for cifar10}, and  \ref{sscc recovery images for cifar10} for CIFAR-10), it is observed that our proposed deep JSCC framework outperforms other benchmark schemes significantly in terms of data recovery task, which shows the efficiency of our proposed framework. Finally, from Figs. \ref{sscc recovery images for mnist} and \ref{sscc recovery images for cifar10}, it is observed that the SSCC scheme fails to recover the transmitted image under the considered scenario with low SNR (i.e., $0~{\rm dB}$). This is because, under such circumstance, the lower bound of transmission rate $R_{\min}$ for JPEG2000 source coding exceeds the maximum rate $R_{\max}$ in \eqref{max rate}, which shows the superiority of the proposed deep JSCC framework beyond the traditional SSCC scheme, especially under bad channel conditions (in low SNR regimes).

\begin{figure}
\centering
\subfigure[MNIST]{
\centering
\includegraphics[width=0.5\linewidth]{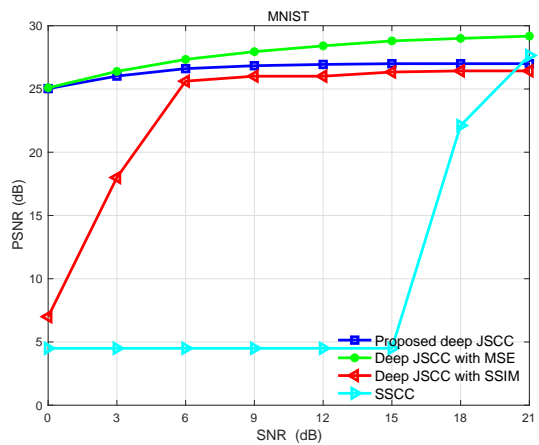}
}%
\subfigure[CIFAR-10]{
\centering
\includegraphics[width=0.5\linewidth]{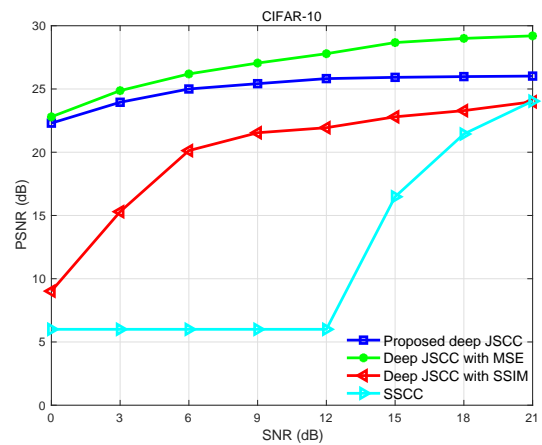}
}%
\caption{The PSNR of our proposed deep JSCC framework and benchmark schemes on MNIST and CIFAR-10 datasets w.r.t. the SNR under AWGN channels.}
\label{PSNR}
\vspace{-20pt}
\end{figure}

Furthermore, we show the performance of our proposed deep JSCC framework on performing image recovery and classification tasks simultaneously as compared to other benchmark schemes. Particularly, we first focus on the image recovery performance in terms of PSNR, as shown in Fig. \ref{PSNR}. First, it is observed from Fig. \ref{PSNR} that the deep JSCC with MSE serves the performance upper bound for the proposed deep JSCC framework, as PSNR is proportional to MSE from \eqref{PSNR equation}. Second, it is observed that our proposed deep JSCC
framework significantly outperforms the SSCC scheme, especially in low SNR regimes, while it has comparable performance with the capacity-achieving SSCC scheme at high SNR values. Third, it is observed that the proposed deep JSCC method does not suffer from the “cliff effect” phenomenon observed from the SSCC scheme.

\begin{figure}
\centering
\subfigure[MNIST]{
\centering
\includegraphics[width=0.5\linewidth]{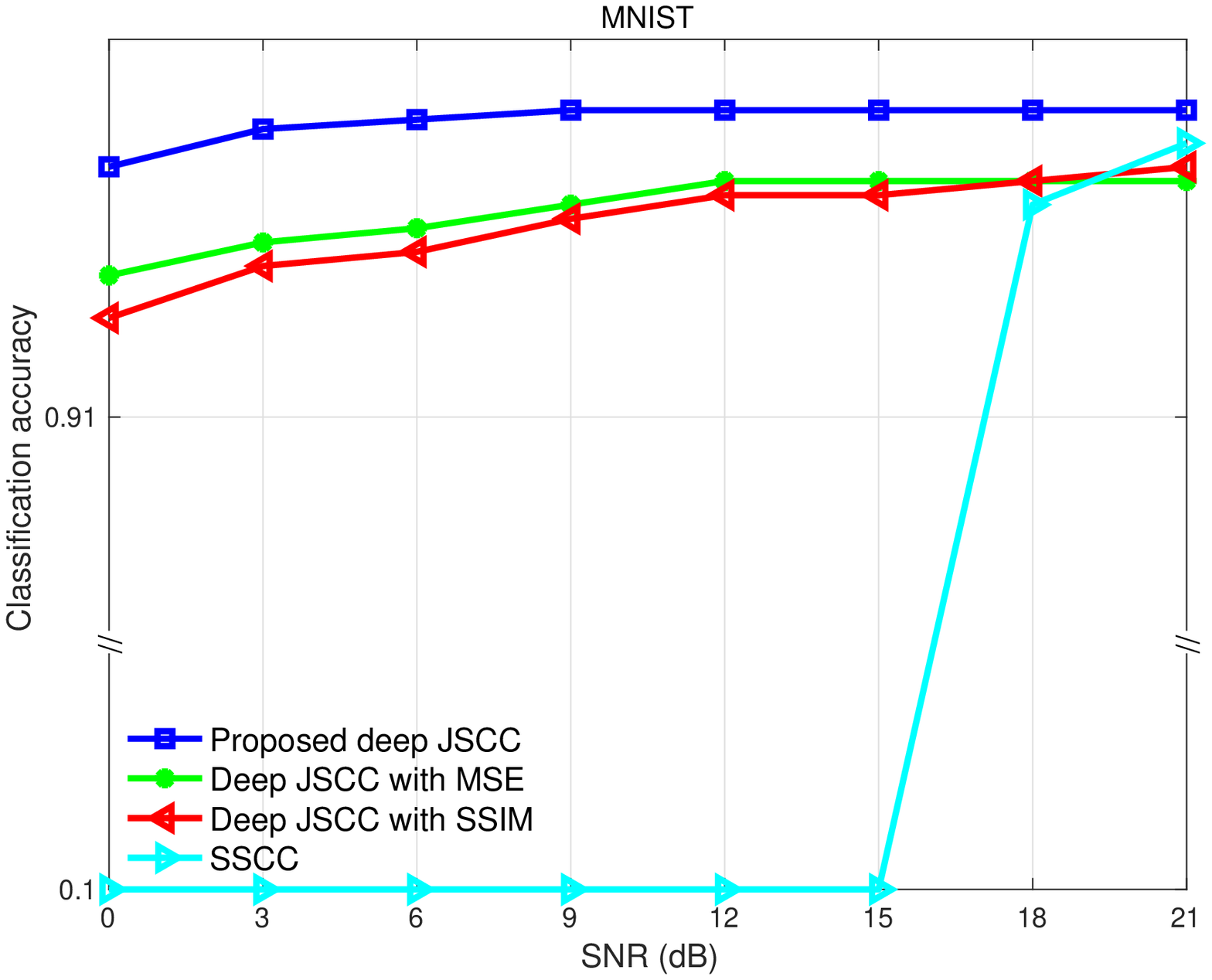}
}%
\subfigure[CIFAR-10]{
\centering
\includegraphics[width=0.5\linewidth]{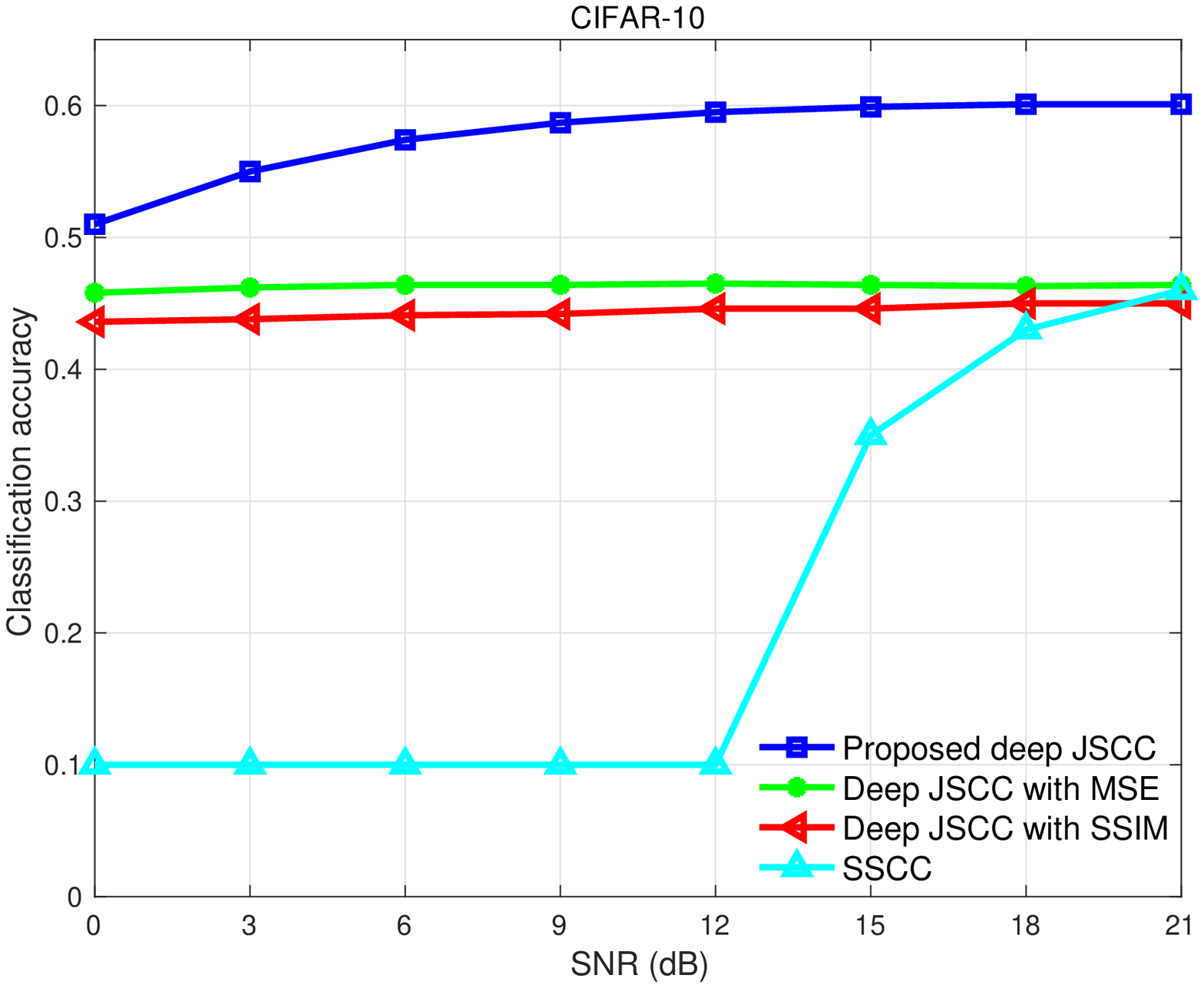}
}%
\caption{Classification accuracy of our proposed deep JSCC framework and the benchmark schemes on MNIST and CIFAR-10 datasets w.r.t. the channel SNR under AWGN channels.}
\label{accuracy}
\vspace{-20pt}
\end{figure}

Finally, we focus on the performance of classification task execution in terms of the classification accuracy when performing image recovery and classification tasks simultaneously, as shown in Fig. \ref{accuracy}. It is observed that our proposed method significantly outperforms other benchmark schemes, which shows that the proposed  deep JSCC framework is able to extract discriminative features for classification directly in the feature space while  performing data recovery tasks simultaneously to provide multi-task services, while the benchmark schemes fail to capture the inherent structure of the data to support classification, and thus results in worse prediction accuracy.
 
\subsection{Performance Evaluation of the Proposed Gated Deep JSCC Framework}
In this subsection, we present the performance of our proposed gated deep JSCC framework facing variational channel conditions as compared to the following benchmark scheme.
\begin{itemize}
	\item \textbf{Proposed deep JSCC trained on ${\bf 21~{\rm {\bf dB}}}$ or ${\bf -3~{\rm {\bf dB}}}$ (without gated net)}: We first train the proposed deep JSCC framework in Fig. \ref{model} under SNR of $21~{\rm dB}$ (or $-3~{\rm dB}$), then we directly transfer the obtained deep JSCC frameworks to variational channel conditions.
\end{itemize}

\begin{figure}
	\centering
	\subfigure[MNIST]{
	\centering
	\includegraphics[width=0.5\linewidth]{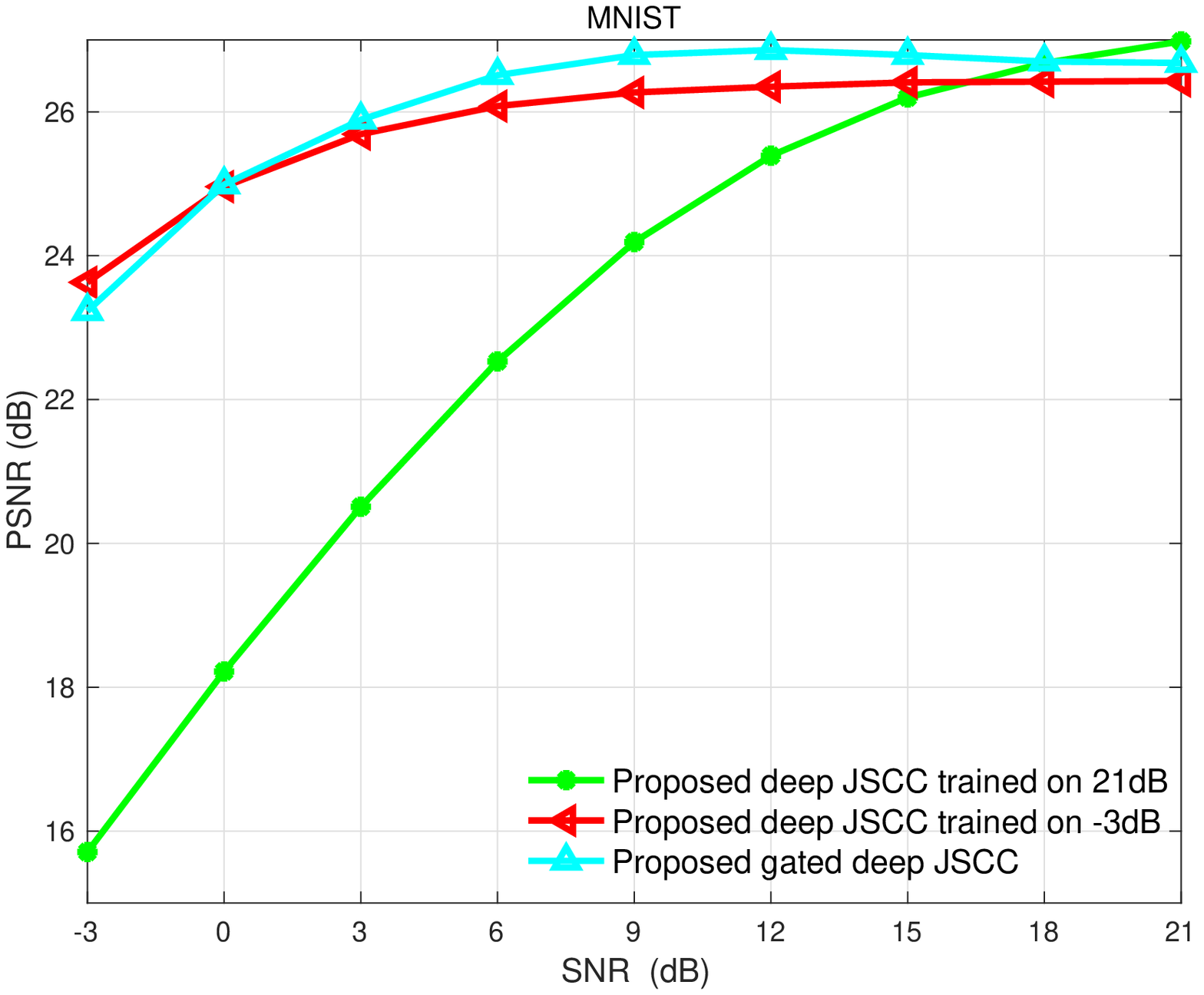}
	}%
	\subfigure[CIFAR-10]{
	\centering
	\includegraphics[width=0.5\linewidth]{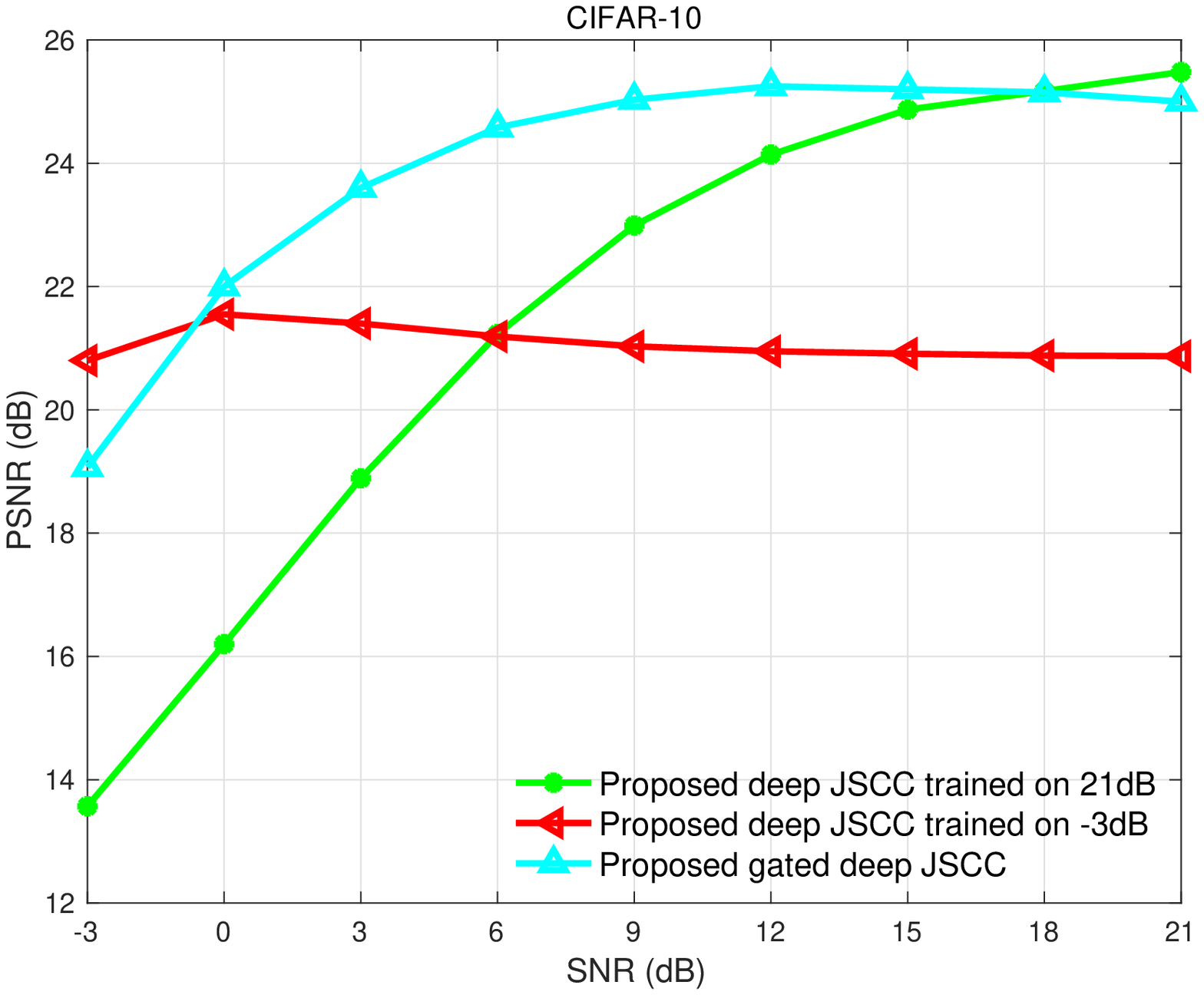}
	}%
	\caption{Performance comparison of our proposed gated deep JSCC framework and benchmark schemes on MNIST and CIFAR-10 datasets in terms of PSNR under variational channel conditions.}
	\label{var_psnr}
	\vspace{-20pt}
\end{figure}

\begin{figure}
	\centering
	\subfigure[MNIST]{
	\centering
	\includegraphics[width=0.5\linewidth]{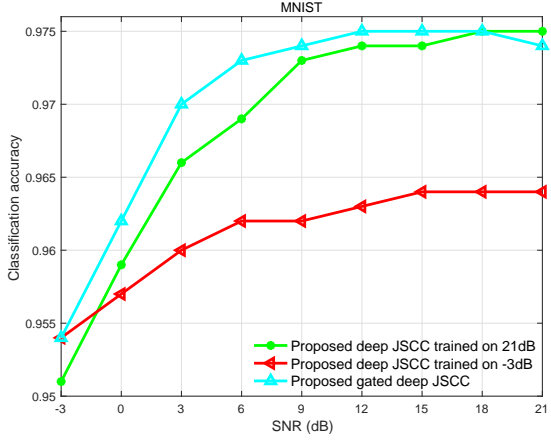}
	}%
	\subfigure[CIFAR-10]{
	\centering
	\includegraphics[width=0.5\linewidth]{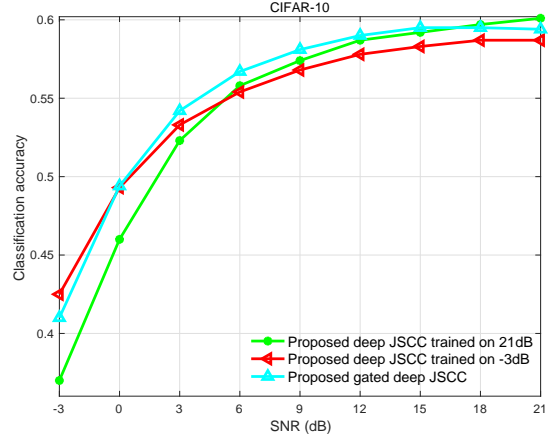}
	}%
	\caption{Performance comparison of our proposed gated deep JSCC framework and benchmark schemes on MNIST and CIFAR-10 datasets in terms of classification accuracy under variational channel conditions.}
	\label{var_acc}
	\vspace{-20pt}
\end{figure}

\begin{figure}
	\centering
	\subfigure[MNIST]{
	\centering
	\includegraphics[width=0.5\linewidth]{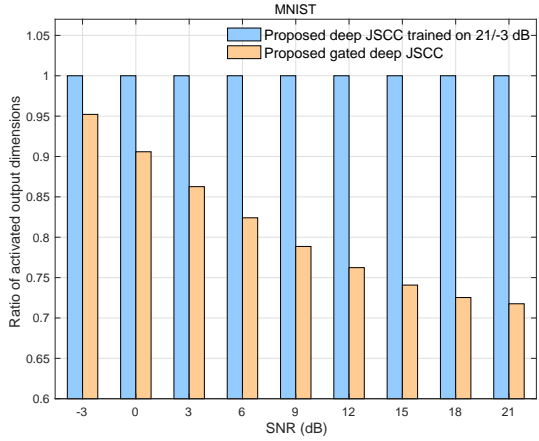}
	}%
	\subfigure[CIFAR-10]{
	\centering
	\includegraphics[width=0.5\linewidth]{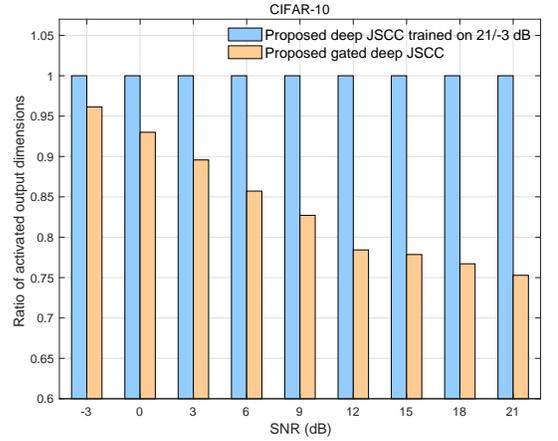}
	}%
	\caption{The ratio of activated output dimensions under variational channel conditions.}
	\label{var_delay}
	\vspace{-20pt}
\end{figure}

Figs. \ref{var_psnr}, \ref{var_acc}, and \ref{var_delay} show the performance of our proposed gated deep JSCC framework and benchmark schemes in terms of PSNR, classification accuracy, and the ratio of activated output dimensions facing variational channel conditions. First, for PSNR and classification accuracy, it is observed from Figs. \ref{var_psnr} and \ref{var_acc} that the models trained on $21~{\rm dB}$ and $-3~{\rm dB}$ merely perform well on high SNR and low SNR regimes, respectively. However, the proposed gated deep JSCC model trained with domain randomization not only shows competitive performance as the benchmark schemes in the high or low SNR regions where they trained on, but also outperforms the benchmark schemes in most of the SNR regimes. Nevertheless, for the ratio of activated output dimensions, it is observed from Fig. \ref{var_delay} that the proposed gated deep JSCC framework enjoys the benefits of lower ratio of activated output dimensions with dynamic feature activation in variational channels (for instance, $4.8\%$ and $28.3\%$ transmission costs are reduced for MNIST image transmission under SNR of $-3~{\rm dB}$ and $21~{\rm dB}$ respectively), compared with the models trained on deterministic channel conditions with fixed output dimensions without significant performance loss.

\subsection{Effects of Label Corruption}
In this subsection, we reveal the benefits of performing classification task directly in the feature space via coding rate reduction maximization in \eqref{rate reduction} in terms of defending label corruption (mislabeling). For label corruption, we randomly mislabel certain ratio of samples in the training set, and then utilize the mislabeled data to train the DNNs to perform classification task. It is worth noting that we neglect the part of data recovery of the model in order to concentrate on the classification performance comparison in this part of experiments. Moreover, we utilize the traditional label-fitting loss function for classification task, i.e., cross-entropy loss, as the benchmark scheme,
\begin{itemize}
	\item {\bf Cross-entropy minimization}: We train the whole network via cross-entropy minimization \cite{IGoodfellow2016}, which is defined as 
	\begin{align}
		\mathrm{Loss}_{\rm cross-entropy}(\omega, \theta)&=-\frac{1}{N}\sum\nolimits_{n = 1}^N {\bar{\boldsymbol{z}}^{\rm T}_n}{\rm log}(\boldsymbol{z}_n)\nonumber \\
		&=-\frac{1}{N}\sum\nolimits_{n = 1}^N {\bar{\boldsymbol{z}}^{\rm T}_n}{\rm log}\bigg(d\Big(p \big({e{(\boldsymbol{s}_n,\theta) }}\big) +\hat{\boldsymbol{n}},\omega\Big)\bigg),
	\end{align}
	where $\bar{\boldsymbol{z}}_n$ denotes the one-hot ground-truth label vector of sample $n$, $\boldsymbol{z}_n$ denotes the prediction result (the output of the DNN) of sample $n$. It  is worth noting that  we modify the output layer at the receiver side for label prediction, to make the output dimension same as the number of classes of the datasets. With the DNN trained via cross-entropy minimization, we perform the classification task and calculate the classification accuracy based on the output prediction results.
\end{itemize}

\begin{figure}
	\centering
	\subfigure[MNIST]{
	\centering
	\includegraphics[width=0.5\linewidth]{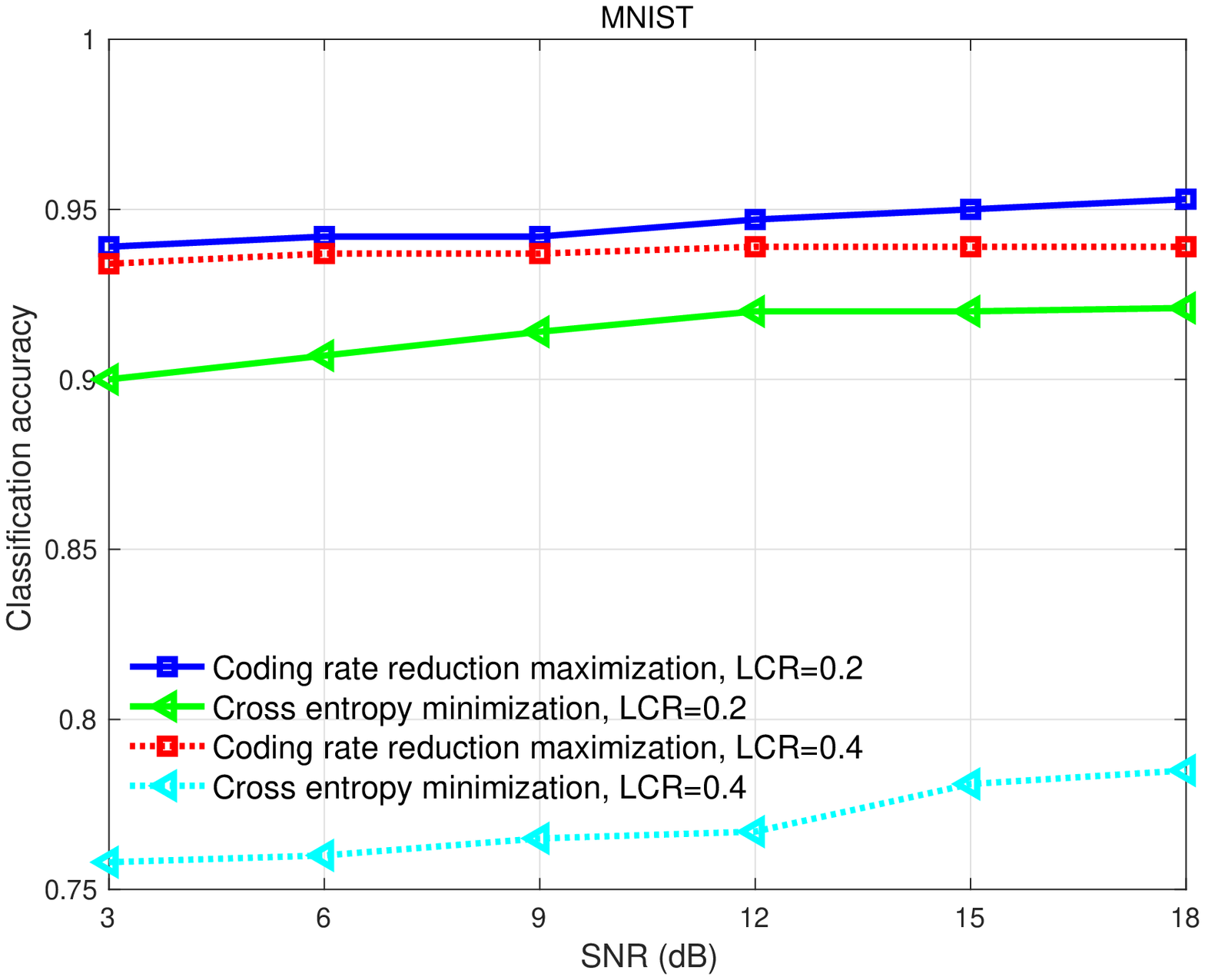}
	}%
	\subfigure[CIFAR-10]{
	\centering
	\includegraphics[width=0.5\linewidth]{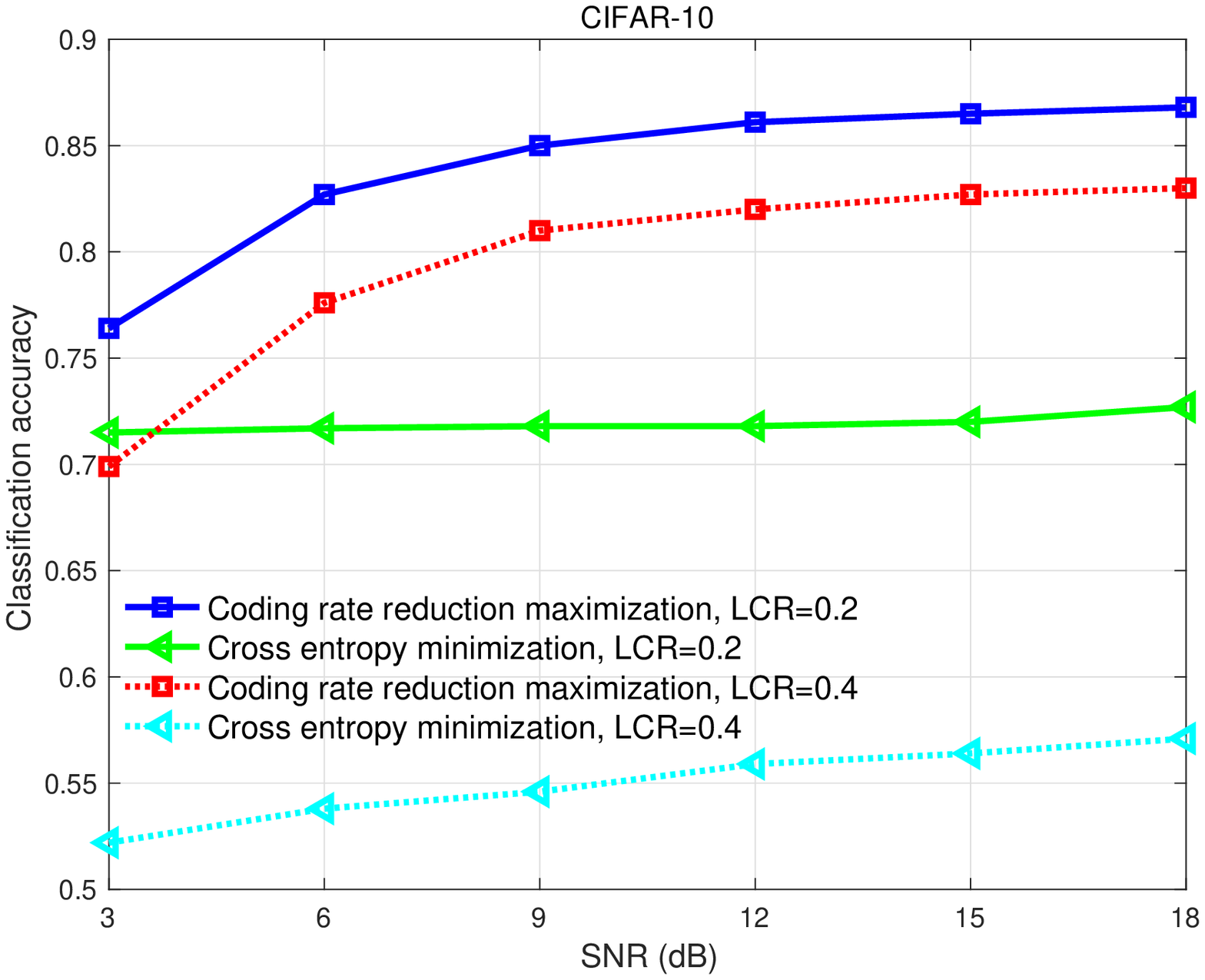}
	}%
	\vspace{-10pt}
	\caption{Performance comparison between utilizing coding rate reduction maximization and cross-entropy minimization in terms of classification accuracy considering label corruption under different LCRs.}
	\label{label_corruption}
	\vspace{-20pt}
\end{figure}

Fig. \ref{label_corruption} shows the performance comparison between the considered coding rate reduction maximization and the cross-entropy minimization to train the DNN for classification task under different label corruption rates (LCRs). It is observed that executing classification directly in the feature space via the considered coding rate reduction maximization outperforms fitting labels via cross-entropy minimization  under label corruption circumstances.  This shows that cross-entropy minimization suffers from mislabeling more easily while coding rate reduction maximization is more robust than cross-entropy minimization, especially with higher LCR. This is because maximizing coding rate reduction could potentially exploit the inherent structure of the data to defend label corruption to some extend, which depends less on the label information as compared with the label-fitting cross-entropy method. 

\section{Conclusions}
Semantic communication has drawn sparkled research interests recently for fully unlock the potential of future wireless networks in terms of supporting AI applications. In this paper, we considered to support multi-task services by performing image recovery and classification task execution at the same time via designing a deep JSCC framework. Specifically, to train the proposed deep JSCC framework, we designed a unified loss function: on one hand, we maximized the coding rate reduction to learn discriminative features for downstream classification tasks; On the other hand, we minimized the MSE of the original and recovered images to learn informative features for data recovery. Furthermore, to further reduce communication overhead and improve the robustness of the model against variational channel conditions, we proposed a gated deep JSCC framework training via domain randomization. Finally, we conducted various experiments to reveal the effectiveness of our proposed designs.

\end{document}